\documentclass[letterpaper,11pt,twoside,fleqn]{book}

\usepackage{mathptmx}

\usepackage{setspace}

\usepackage[verbose,letterpaper,portrait,twoside,dvips,heightrounded]{geometry}

\usepackage[multiple,stable]{footmisc}

\usepackage[small,bf]{caption2}

\setcaptionmargin{1in}

\usepackage[footnotesize,bf,FIGBOTCAP]{subfigure}

\usepackage{fancyhdr}

\fancypagestyle{plain}{%
  \fancyhf{}
  
  \fancyhf[FLE,FRO]{\hfill\slshape\thepage}}

\makeatletter
\renewcommand{\chaptermark}[1]{\markboth{\@chapapp\ \thechapter.\ \ #1}{}}
\makeatother

\usepackage{sectsty}

\subsubsectionfont{\vspace{5pt}\slshape\mdseries}

\usepackage[dvips]{color}
\usepackage[grey]{quotchapdan}

\definecolor{chaptergrey}{rgb}{0.1,0.1,0.1}

\renewcommand{\numberalign}{\raggedleft}
\renewcommand{\textalign}{\raggedleft}

\usepackage[subfigure]{tocloft}

\tocloftpagestyle{plain}

\makeatletter
\renewcommand{\cftmarktoc}{\@mkboth{\contentsname}{\contentsname}}
\renewcommand{\cftmarklof}{\@mkboth{\listfigurename}{\listfigurename}}
\renewcommand{\cftmarklot}{\@mkboth{\listtablename}{\listtablename}}
\makeatother

\renewcommand{\contentsname}{Contents}

\renewcommand{\listfigurename}{List of Figures}

\renewcommand{\listtablename}{List of Tables}

\usepackage{tocbibind}

\makeatletter
\renewcommand\@makeschapterhead[1]{%
  {\parindent \z@ \centering
    \normalfont
    \interlinepenalty\@M
    \huge \bfseries \sc #1\par\nobreak
    \vskip 10\p@
  }}
\makeatother

\usepackage[toc]{appendix}

\renewcommand{\appendixtocname}{\Large\bfseries\textsc{Appendices}}

\noappendicestocpagenum

\renewcommand{\appendixpagename}{\scshape{\mdseries{Appendices}}}

\renewcommand{\appendixname}{Appendix}

\usepackage{amsmath,amssymb}

\allowdisplaybreaks

\makeatletter
\@ifundefined{mathindent}{\relax}{\setlength{\mathindent}{0pt}}
\makeatother

\usepackage[dvips,final]{graphicx}
\usepackage{epsfig,rotating}

\graphicspath{{c2jets/}{c3sph/}{c45spmhd/}}

\usepackage{natbib}

\usepackage{bibentry}

\usepackage{epigraph}

\renewcommand{\epigraphsize}{\small}

\newlength{\myepigraphwidthfrontend}
\renewcommand{\textflush}{flushepinormal}

\renewcommand{\epigraphflush}{flushleft}

\renewcommand{\sourceflush}{flushright}

\usepackage{array}

\usepackage{booktabs}

\def\sma{\sum_a m_a}
\def\smb{\sum_b m_b}
\def\smc{\sum_c m_c}
\def\va{\mathbf{v}_a}

\def\gawab{\nabla_a W_{ab}}
\def\runit{\hat{\mathbf{r}}_{ab}}

\def\bv{\mathbf{v}}

\def\br{\mathbf{r}}

\def\sma{\sum_a m_a}
\def\smb{\sum_b m_b}
\def\smc{\sum_c m_c}
\def\gwab{\nabla_a W_{ab}}

\newcommand{\pder}[2]{\ensuremath{\frac{\partial #1}{\partial #2}}}

\def\rv{\mathrm{v}}

\makeatletter
\renewcommand*{\cleardoublepage}{%
  \clearpage
  \if@twoside
    \ifodd\c@page
    \else
      \thispagestyle{empty}     
      \hbox{}\newpage
      \if@twocolumn
        \hbox{}\newpage
      \fi
    \fi
  \fi
}
\makeatother
  
\begin{document}



\mainmatter

\setcounter{equation}{0}

\include{c1}

\include{c2}

\chapter{Smoothed Particle Hydrodynamics}
\label{chap:sph}
\epigraphhead[70]{\epigraph{%
    ``I went on to test the program in every way I could devise.  I strained
it to expose its weaknesses.  I ran it for high-mass stars and low-mass
stars, for stars born exceedingly hot and those born relatively cold.
I ran it assuming the superfluid currents beneath the crust to be
absent -- not because I wanted to know the answer, but because I had
developed an intuitive feel for the answer in this particular case.
Finally I got a run in which the computer showed the pulsar's
temperature to be less than absolute zero.  I had found an error.  I
chased down the error and fixed it.  Now I had improved the program to
the point where it would not run at all.''
  }{%
    \textit{Frozen Star: Of Pulsars, Black Holes and the Fate of Stars} \\
    \textsc{George Greenstein}}}


\section{Introduction}
 The standard approach to solving the equations of fluid dynamics numerically
is to define fluid quantities on a regular spatial grid, computing derivatives
using finite difference or finite volume schemes. This is an extremely well studied approach and most
`state of the art' methods for fluid dynamics have been developed in this manner.
In astrophysical fluid dynamics problems frequently involve changes in spatial,
temporal and density scales over many orders of magnitude. Thus, adaptivity is
an essential ingredient which is absent from a fixed-grid approach. Progress in
this area has been rapid in recent years with the development of procedures for
adaptive mesh refinement (AMR). The implementation of such procedures is far
from trivial, although the availability of libraries and toolkits for
grid-based codes eases this burden somewhat. However, a further constraint is
that astrophysical problems are frequently asymmetric which can result in
substantial numerical diffusion when solving on (fixed or adaptive) Cartesian
grids. Other approaches to this problem are to use unstructured grids (where typically the grid is
reconstructed at each new timestep) or Lagrangian grid methods, where the grid shape
deforms according to the flow pattern. 

 An alternative to all of these methods is to remove the spatial grid entirely, 
resulting in methods which are inherently adaptive. In
this approach fluid quantities are carried by a set of moving interpolation points
which follow the fluid motion. Since each point carries a fixed mass, the
interpolation points are referred to as `particles'. Derivatives are evaluated either
by interpolation over neighbouring particles (referred to as particle methods), or via a hybrid
approach by interpolation to an overlaid grid (referred to as particle-mesh methods,
typified by the particle-in-cell (PIC) method used extensively in plasma
physics.

Smoothed Particle Hydrodynamics (SPH) is a particle
method introduced by \citet{lucy77} and \citet{gm77}. It has found widespread use in
astrophysics due to its ability to tackle a wide range of problems
involving complex, asymmetric phenomena with relative ease. Since these features are
highly desirable in many non-astrophysical applications, it is unsurprising that
SPH is currently finding many applications in other fields such as geophysics and
engineering (and even film-making\footnote{for example many of the graphics involving
fluids in the film `Tomb Raider' were computed using SPH}). 

 The advantages of SPH over standard grid based approaches can be summarised as
follows: Firstly, SPH is conceptually both simple and beautiful. All of the equations can be
derived self-consistently from physical principles with a few basic assumptions.
As a result complex physics is relatively simple to incorporate. Its simplicity
means that for the user it is a very intuitive numerical method which lends
itself easily to problem-specific modifications. Secondly, adaptivity is a
built-in feature. The Lagrangian nature of the method means that changes in
density and flow morphology are automatically accounted for without the need for
mesh refinement or other complicated procedures. As a result of its adaptivity,
SPH is also very efficient in that resolution is concentrated on regions of high density, whilst
computational effort is not wasted on empty regions of space. Thirdly, free
boundaries, common in astrophysical problems, are simple and natural in SPH but
often present difficulties for grid-based codes (such as spurious heating from
the interaction with a low density surrounding medium). This means that no
portions of fluid can be lost from the simulation, unlike in a grid based code
where fluid which has left the grid cannot return (this has been dubbed the
`Columbus effect' by Melvyn Davies, since fluid can fall off the edge of the
world). Fourthly, a significant advantage in an astrophysical context is that
SPH couples naturally with widely used N-Body codes and techniques, for which
there exists a vast amount of literature. Finally (although perhaps many more
advantages could be given) visualisation and analysis is also somewhat easier with Lagrangian techniques, since
it is a simple matter to track and visualise portions of the flow.

 SPH also has a number of disadvantages
when compared to finite difference codes. The first of these is that, unlike
grid-based codes, SPH involves the additional computational cost of constructing the
neighbour lists. This is offset somewhat in that N-Body techniques used to
calculate the gravitational force (namely via tree-codes) can also be used in
constructing the neighbour lists. Secondly, SPH suffers from a lack of algorithm development,
since a vast amount of research effort is focussed on finite difference
or finite volume techniques. This often means that such techniques, although often applicable in
an SPH context, can be slow to filter into mainstream use. Thirdly, although not
a disadvantage as such but a point which is often overlooked, is that the setup
of initial conditions is often more complicated and requires much greater care. Since
particles can be laid down in an infinite variety of ways, choosing an
appropriate setup for a given problem requires some experience and usually some
experimentation. Inappropriate particle setups can lead to poorer simulation
results than might otherwise be expected (we give some examples
of this in \S\ref{sec:shearflows}). Finally, in the case of magnetohydrodynamics and other problems involving anisotropic stresses
(as we will discuss in chapter \ref{chap:spmhd}), numerical stability can
become an issue which must be dealt with appropriately.

 In this chapter we provide an overview of the SPH method, including several
improvements to the basic method which have been made since the review
article of \citet{monaghan92} was published (such as improvements in
shock-capturing techniques and the treatment of terms related to the use of a variable smoothing
length). In particular we focus on those aspects of the algorithm that are
relevant in an MHD context. The chapter is organised as follows: In section \S\ref{sec:basics} we present
the basic formalisms inherent to SPH; in \S\ref{sec:fluideqs} we derive the
SPH equations for compressible hydrodynamics using a variational principle. Formulations of dissipative terms used
to capture shocks are presented and discussed in \S\ref{sec:av}. In
\S\ref{sec:gradh} we discuss the incorporation of terms relating to the spatial 
variation of the smoothing length and in \S\ref{sec:altforms} alternative formulations of
SPH are examined within the variational framework. Timestepping is discussed in
\S\ref{sec:timestep}. Finally, we present numerical
tests in \S\ref{sec:hydrotests} in support of the previous sections and as
preliminaries for the MHD tests described in Chapters \ref{chap:spmhd} and
\ref{chap:divb}.

\section{Basic formalisms}
\label{sec:basics}

\subsection{Interpolant}
 The basis of the SPH approach is given as follows \citep{monaghan92}. We begin with the trivial
identity\footnote{It is interesting to note that this equation, with $A = \rho$ is used to define
the density of the fluid in terms of the Lagrangian co-ordinates  in the Hamiltonian description of
the ideal fluid \citep[eq. (94) in][]{morrison98}.  Similarly the SPH
equivalent of this expression, (\ref{eq:rhosum}), forms the basis for the Hamiltonian
description of SPH (see \S\ref{sec:sphmom}).}
\begin{equation}
A(\mathbf{r}) = \int A(\mathbf{r'}) \delta(|\mathbf{r-r'}|) d\mathbf{r'},
\end{equation}
where $A$ is any variable defined on the spatial co-ordinates $\mathbf{r}$ and $\delta$ refers to the Dirac delta function. This integral is then
approximated by replacing the delta function with a smoothing kernel $W$ with
characteristic width $h$, such that
\begin{equation}
\lim_{h\to 0} W(\br - \br',h) = \delta (\br - \br'),
\label{eq:wtodelta}
\end{equation}
giving
\begin{equation}
A(\mathbf{r}) = \int A(\mathbf{r'}) W(|\mathbf{r-r'}|, h) d\mathbf{r'} + O(h^2).
\label{eq:intint}
\end{equation}
The kernel function is normalised according to
\begin{equation}
\int W(\br - \br',h) \mathrm{d\br'} = 1.
\label{eq:kernelnorm}
\end{equation}
 Finally the integral (\ref{eq:intint}) is discretised onto a finite set of interpolation points
(the particles) by replacing the integral by a summation and the mass element $\rho dV$ with the
particle mass $m$, ie.
\begin{eqnarray}
A(\mathbf{r}) & = & \int \frac{A(\mathbf{r'})}{\rho(\mathbf{r'})}
W(|\mathbf{r-r'}|, h) \rho(\mathbf{r'})d\mathbf{r'} + O(h^2), \nonumber \\
& \approx & \sum^{N}_{b=1} m_{b} \frac{A_{b}}{\rho_b}
W(|\mathbf{r}-\mathbf{r}_{b}|, h), \label{eq:sumint}
\end{eqnarray}
where the subscript $b$ refers the quantity evaluated at the position of particle $b$.
This `summation interpolant' is the basis of all SPH formalisms. The errors
introduced in this step are discussed in \S\ref{sec:errors}. Gradient terms may be calculated by taking the analytic derivative of (\ref{eq:sumint}), giving
\begin{eqnarray}
\nabla A(\mathbf{r}) & = & \frac{\partial}{\partial \mathbf{r}} \int
\frac{A(\mathbf{r'})}{\rho(\mathbf{r'})} W(|\mathbf{r-r'}|, h)
\rho(\mathbf{r'})d\mathbf{r'} + O(h^2), \label{eq:gradintint} \\
& \approx & \smb \frac{A_{b}}{\rho_b} \gawab,\label{eq:gradsumint}
\end{eqnarray}
where we have assumed that the gradient is evaluated at another particle $a$ (ie.
$\br = \br_a$), defining $\nabla_{a}\equiv \frac{\partial}{\partial \mathbf{r}_a}$ and $W_{ab}\equiv W(|\mathbf{r}_{a}-\mathbf{r}_{b}|,
h)$. 

\subsection{Errors}
\label{sec:errors}
The errors introduced by the approximation (\ref{eq:intint}) can be estimated by
expanding $A(\br')$ in a Taylor series about $\br$ \citep{benz90,monaghan92}, giving
\begin{eqnarray}
A(\br) & = & \int \left[A(\br) + (\br'-\br)^\alpha \pder{A}{\br^\alpha} + \frac12
(\br'-\br)^\beta (\br'-\br)^\gamma \pder{^2 A}{\br^\beta \partial\br^\gamma} +
\mathcal{O}((\br-\br')^3)\right] W(\vert\br-\br'\vert, h) \mathrm{d\br'}, \nonumber \\
& = & A(\br) + \pder{A}{\br^\alpha} \int (\br'-\br)^\alpha W(r) \mathrm{d\br'} + \frac12
\pder{^2 A}{\br^\beta \partial\br^\gamma} \int (\br'-\br)^\beta (\br'-\br)^\gamma W(r)
\mathrm{d\br'} + \mathcal{O}[(\br'-\br)^3],
\label{eq:kernelerrors}
\end{eqnarray}
where $r\equiv\vert \br'-\br \vert$; $\alpha, \beta$ and $\gamma$ are indices denoting co-ordinate directions (with repeated indices implying a summation) and we have used the normalisation condition (\ref{eq:kernelnorm}). The odd error terms are zero if $W$ is an even function of
$(\br-\br')$ (ie. depending only on its magnitude), which, since $\vert \br-\br'
\vert$ is always less than the smoothing radius ($2h$ in most cases), results in
an approximation to $\mathcal{O}(h^2)$. In principle it is
also possible to construct kernels such that the second moment is also zero,
resulting in errors of $\mathcal{O}(h^4)$ (discussed further in
\S\ref{sec:kernelstability}). The disadvantage of such kernels is
that the kernel function becomes negative in some part of the domain, resulting
in a potentially negative density evaluation. The errors in the summation
interpolant differ slightly since the approximation of integrals by summations
over particles no longer guarantees that these terms integrate
exactly. Starting from the summation interpolant evaluated on particle $a$, we
expand $A_b$ in a Taylor series around $\br_a$, giving
\begin{equation}
\smb \frac{A_b}{\rho_b} W_{ab} = A_a \sum_b \frac{m_b}{\rho_b} W_{ab} + \nabla A_a \cdot \sum_b
\frac{m_b}{\rho_b} (\br_b - \br_a) W_{ab} + \mathcal{O}[(\br_b-\br_a)^2].
\end{equation}
From this we see that the summation interpolation is exact for constant
functions only when the interpolant is normalised by dividing by the
interpolation of unity. In practical calculations the summation interpolant is
only used in the density evaluation (\S\ref{sec:sphcty}), resulting in a slight
error in the density value. More important are the errors resulting from the SPH
evaluation of derivatives, since these are used throughout in the discretisation
of the fluid equations (\S\ref{sec:fluideqs}).

 The errors resulting from the gradient evaluation (\ref{eq:gradintint}) may be
estimated in a similar manner by again expanding $A(\br')$ in a Taylor series about $\br$, giving
\begin{eqnarray}
\nabla A(\mathbf{r}) & = & \int \left[A(\mathbf{r}) + (\br' - \br)^\alpha
\pder{A}{\br^\alpha} + \frac12 (\br'-\br)^\beta (\br'-\br)^\gamma
\pder{^2 A}{\br^\beta \partial \br^\gamma} + \mathcal{O}[(\br-\br')^3] \right] \nabla
W(|\mathbf{r-r'}|, h) d\mathbf{r'}, \nonumber \\ 
& = & A(\mathbf{r}) \int \nabla W d\mathbf{r'} + \pder{A}{\br^\alpha}\int
(\br'-\br)^\alpha \nabla
W \mathrm{d\br'} + \frac12 \pder{^2 A}{\br^\beta \partial\br^\gamma} \int
 (\br'-\br)^\beta (\br'-\br)^\gamma \nabla W \mathrm{d\br'} +
\mathcal{O}[(\br' - \br)^3], \nonumber \\
& = & \nabla A(\mathbf{r}) + \frac12 \pder{^2 A}{\br^\beta \partial \br^\gamma} \int
(\br'-\br)^\beta (\br'-\br)^\gamma \nabla W(r) \mathrm{d\br'} +
\mathcal{O}[(\br'-\br)^3], \label{eq:gradinterrors}
\end{eqnarray}
where we have used the fact that $\int \nabla W d\mathbf{r'} = 0$ for even
kernels, whilst the second term
integrates to unity for even kernels satisfying the normalisation condition
(\ref{eq:kernelnorm}). The resulting errors in the integral interpolant for the
gradient are therefore also of $\mathcal{O}(h^2)$. The errors in the summation
interpolant for the gradient (\ref{eq:gradsumint}) are given by expanding $A_b$ in a Taylor series around $\br_a$, giving
\begin{eqnarray}
\nabla A_a & = & \smb \frac{A_{b}}{\rho_b} \gawab, \nonumber \\
& = & A_a \sum_b \frac{m_b}{\rho_b} \gawab + \pder{A_a}{\br^\alpha}  \sum_b \frac{m_b}{\rho_b} (\br_b - \br_a)^\alpha \gawab \nonumber \\
& & \phantom{A_a \sum_b \frac{m_b}{\rho_b} \gawab} + \frac12 \pder{^2
A_a}{\br^\beta \partial\br^\gamma} \sum_b
\frac{m_b}{\rho_b} (\br_b - \br_a)^\beta (\br_b - \br_a)^\gamma \gawab +
\mathcal{O}[(\br_b-\br_a)^3]. \label{eq:sumgraderrors}
\end{eqnarray}
where the summations represent SPH approximations to the integrals in
the second line of (\ref{eq:gradinterrors}).

\subsection{First derivatives}
\label{sec:gradients}
 From (\ref{eq:sumgraderrors}) we immediately see that a straightforward
improvement to the gradient estimate (\ref{eq:gradsumint}) can be obtained by a simple subtraction of
the first error term (i.e. the term in (\ref{eq:sumgraderrors}) that is present even in the case of a constant function), giving \citep{monaghan92}
\begin{equation}
\nabla A_a = \smb \frac{(A_b - A_a)}{\rho_b} \gawab \label{eq:bettergrad}, 
\end{equation}
which is an SPH estimate of
\begin{equation}
\nabla A(\mathbf{r}) = \nabla A - A (\nabla 1).
\end{equation}
Since the first error term in (\ref{eq:sumgraderrors}) is removed, the
interpolation is exact
for constant functions and indeed this is obvious from the form of
(\ref{eq:bettergrad}).
The interpolation can be made exact for linear functions
by dividing by the summation multiplying the first derivative term in
(\ref{eq:sumgraderrors}), ie.
\begin{equation}
\pder{A_a}{\br^\alpha} = \chi_{\alpha\beta} \sum_b \frac{m_b}{\rho_b}(A_b - A_a)
\nabla^\beta W_{ab}, \hspace{1cm} \chi_{\alpha\beta} = \left[
\sum_b\frac{m_b}{\rho_b} (\br_b -
\br_a)^\alpha \nabla^\beta W_{ab} \right]^{-1}.  
\label{eq:exactlinear}
\end{equation}
where $\nabla^\beta \equiv \partial / \partial \br^\beta$. This normalisation is somewhat cumbersome in practice, since $\chi$ is a matrix quantity, requiring considerable
extra storage (in three dimensions this means storing $3\times 3 = 9$ extra quantities
for each particle) and also since calculation of this term requires prior knowledge of
the density. However, for some applications of SPH (e.g. solid mechanics) it
is desirable to do so in order to retain angular momentum conservation in the
presence of anisotropic forces \citep{bl99}.

 A similar interpolant for the gradient follows by using
\begin{eqnarray}
\nabla A & = & \frac{1}{\rho} [A\nabla \rho -\nabla(\rho A)] \\
 & \approx & \frac{1}{\rho_a} \smb (A_b - A_a) \gawab, \label{eq:bettergradusual}
\end{eqnarray}
which again is exact for a constant $A$. Expanding $A_b$ in a Taylor series, we
see that in this case the interpolation of a linear function
can be made exact using
\begin{equation}
\pder{A_a}{\br^\alpha} = \chi_{\alpha\beta} \smb (A_b - A_a)
\nabla^\beta W_{ab}, \hspace{1cm} \chi_{\alpha\beta} = \left[
\smb (\br_b -
\br_a)^\alpha \nabla^\beta W_{ab} \right]^{-1}.  
\end{equation}
which has some advantages over (\ref{eq:exactlinear}) in that it can be computed without prior
knowledge of the density.

 An alternative gradient interpolant is given by
\begin{eqnarray}
\nabla A(\mathbf{r}) & = & \rho \left[\frac{A}{\rho^2}\nabla\rho +
\nabla\left(\frac{A}{\rho}\right) \right] \nonumber \\
& \approx & \rho_a \smb \left( \frac{A_a}{\rho_a^2} +
\frac{A_b}{\rho_b^2}\right)\gawab \label{eq:symmetrisedgrad}
\end{eqnarray}
which is commonly used in the SPH evaluation of the pressure gradient since it
guarantees conservation of momentum by the pairwise symmetry in the gradient
term. It is also the formulation of the pressure gradient which follows
naturally in the derivation of the SPH equations from a variational principle
(\S\ref{sec:sphmom}). Expanding $A_b$ in a Taylor series about $\br_a$ we have
\begin{eqnarray}
\smb \left( \frac{A_a}{\rho_a^2} +
\frac{A_b}{\rho_b^2}\right)\gawab & = & A_a \smb \left( \frac{1}{\rho_a^2} +
\frac{1}{\rho_b^2}\right)\gawab + \pder{A_a}{\br^\alpha} \sum_b \frac{m_b}{\rho_b^2} (\br_b -
\br_a)^\alpha \gawab \nonumber \\
& & + \frac12 \pder{^2 A_a}{\br^\beta \partial\br^\gamma} \sum_b \frac{m_b}{\rho_b^2} (\br_b -
\br_a)^\beta (\br_b - \br_a)^\gamma \gawab + \mathcal{O}[(\br_b-\br_a)^3] \label{eq:aboringline}
\end{eqnarray}
from which we see that for a constant function the error is governed by the extent to which
\begin{equation}
\smb \left( \frac{1}{\rho_a^2} + \frac{1}{\rho_b^2}\right)\gawab \approx 0.
\end{equation}
Although a simple subtraction of the first term in (\ref{eq:aboringline})
from (\ref{eq:symmetrisedgrad}) eliminates this error, the symmetry in the gradient necessary for the
conservation of momentum is lost by doing so. Retaining the exact conservation of
momentum therefore requires that such error terms are not eliminated. In
applications of SPH employing anisotropic forces (such in the MHD case), these error terms can be
sufficient to cause numerical instabilities (\S\ref{sec:mhdstability}).

 Derivatives of vector quantities follow in a similar manner. For
example the divergence of a vector quantity $\bv$ can be estimated using
\begin{equation}
(\nabla\cdot\bv)_a \approx -\frac{1}{\rho_a} \smb (\bv_a - \bv_b)\cdot \gawab, 
\label{eq:sphdivergence}
\end{equation}
or
\begin{equation}
(\nabla\cdot\bv)_a \approx \rho_a \smb  \left( \frac{\bv_a}{\rho_a^2} +
\frac{\bv_b}{\rho_b^2}\right)\cdot \gawab, 
\label{eq:sphdivergence2}
\end{equation}
whilst the curl is given by (e.g.)
\begin{equation}
(\nabla \times \bv)_a \approx -\frac{1}{\rho_a} \smb (\bv_a - \bv_b)\times \gawab.
\label{eq:sphcurl}
\end{equation}

\subsection{Second derivatives}
\label{sec:2ndderiv}
Second derivatives are slightly more complicated since for kernels with compact
support a straightforward estimation using the second derivative of the
kernel proves to be very noisy and sensitive to particle disorder. For this reason it is better to use
approximations of the second derivative which utilise only the first derivative
of the kernel \citep{brookshaw85,monaghan92}. For a scalar quantity the second derivative may be estimated
using the integral approximation
\begin{equation}
\nabla^2 A(\br) \approx 2 \int [A(\br) - A(\br')] \frac{(\br - \br') \cdot
\nabla W(\br)}{\vert \br - \br' \vert^2} \mathrm{d\br'},
\label{eq:intlaplacian}
\end{equation}
giving the SPH Laplacian
\begin{equation}
(\nabla^2 A)_a \approx 2 \smb \frac{(A_a - A_b)}{\rho_b } \frac{\br_{ab}\cdot
\gawab}{\br_{ab}^2},
\label{eq:sphlaplacian}
\end{equation}
where $\br_{ab} \equiv \br_a - \br_b$. This formalism is commonly used for heat conduction in SPH (e.g.
\citealt{brookshaw85,cm99} and more recently \citealt{jsd04}).
The integral approximation (\ref{eq:intlaplacian}) can be derived by expanding
$A(\br')$ to second order in a Taylor series about $\br$, giving
\begin{equation}
A(\br) - A(\br') = (\br - \br')^\alpha \pder{A}{\br^\alpha} + \frac12 (\br - \br')^\alpha(\br -
\br')^\beta \pder{^2 A}{\br^\alpha \partial \br^\beta} + \mathcal{O}[(\br - \br')^3].
\end{equation}
Expanding this expression into (\ref{eq:intlaplacian}), the integral is given by
\begin{equation}
\pder{A}{\br^\alpha} \int (\br - \br')^\alpha \frac{(\br - \br') \cdot \nabla W(\br)}{\vert \br
- \br' \vert^2}\mathrm{d\br'} + \frac12 \pder{^2 A}{\br^\alpha \partial \br^\beta} \int (\br -
\br')^\alpha (\br - \br')^\beta \frac{(\br - \br') \cdot \nabla W(\br)}{\vert \br
- \br' \vert^2}\mathrm{d\br'}.
\end{equation}
The first integral is zero for spherically symmetric kernels, whilst the second term
integrates to a delta function, giving $\nabla^2 A$. A generalisation of
(\ref{eq:sphlaplacian}) is derived for vector quantities by \citet{er03}. In
three dimensions the integral approximation is given by
\begin{equation}
\pder{^2 \bv}{\br^\alpha \partial \br^\beta} \approx \int [\bv(\br) - \bv(\br')]
\left[ 5(\br - \br')^\alpha (\br - \br')^\beta - \delta^{\alpha\beta} \right]
\frac{(\br - \br') \cdot \nabla W(\br)}{\vert \br - \br' \vert^2} \mathrm{d\br'},
\label{eq:intveclaplacian}
\end{equation}
which in SPH form becomes
\begin{equation}
\left( \pder{^2 \bv}{\br^\alpha \partial \br^\beta} \right)_a \approx \smb \frac{(\bv_a -
\bv_b)}{\rho_b } \left[5 \br_{ab}^\alpha \br_{ab}^\beta - \delta^{\alpha\beta} \right] \frac{\br_{ab}\cdot
\gawab}{\br_{ab}^2}.
\label{eq:sphveclaplacian}
\end{equation}

\subsection{Smoothing kernels}
\label{sec:kernels}
 The smoothing kernel $W$ must by definition satisfy the requirement that it
tends to a delta function as the smoothing length $h$ tends to zero (\ref{eq:wtodelta})
and the normalisation condition (\ref{eq:kernelnorm}). In addition the kernel is usually chosen to be an even function of $r$ to cancel the first
error term in (\ref{eq:kernelerrors}) and may therefore be written in the form
\begin{equation}
W(r,h) = \frac{\sigma}{h^\nu} f\left(\frac{r}{h}\right),
\label{eq:evenkernels}
\end{equation}
where $r\equiv \vert \br - \br' \vert$ and $\nu$ is the number of spatial
dimensions. Written in this form the normalisation condition
(\ref{eq:kernelnorm}) becomes
\begin{equation}
\sigma \int f(q) \mathrm{dV} = 1,
\label{eq:kernelnormq}
\end{equation}
where $q=r/h$ and the volume element $dV = dq, 2\pi q dq$ or $4\pi q^2 dq$ in one, two and three
dimensions. The simplest kernel with this property is the Gaussian
\begin{equation}
W(r,h) = \frac{\sigma}{h^\nu} e^{-q^2},
\label{eq:Gaussian}
\end{equation}
where $q = r/h$ and $\sigma = [1/\sqrt{\pi},1/\pi,1/(\pi\sqrt{\pi})]$ in [1,2,3] dimensions. This has the advantage that the spatial derivative is
infinitely smooth (differentiable) and therefore exhibits good stability
properties (Figure \ref{fig:kernelstability}). For practical applications,
however, using a Gaussian kernel has the immediate disadvantage that the
interpolation spans the entire spatial domain (with computational cost of $\mathcal{O}(N^2)$), despite the
fact that the relative contribution from neighbouring particles quickly become
negligible with increasing distance. For this reason it is far more efficient to
use kernels with finite extent (ie. having compact support), reducing the calculation to a sum over closely
neighbouring particles which dramatically reduces the cost to
$\mathcal{O}(nN)$ where $n$ is the number of contributing neighbours (although there
is also the additional cost of finding the neighbouring particles). Kernels which are similar to the Gaussian in shape
generally give the best performance \citep[see, e.g.][]{fq96}. Of these the most commonly used kernel is that based on cubic splines \citep{ml85}, given by
\begin{equation}
f(q) = \sigma \left\{ \begin{array}{ll}
1 - \frac{3}{2}q^2 + \frac{3}{4}q^3, & 0 \le q < 1; \\
\frac{1}{4}(2-q)^3, & 1 \le q < 2; \\
0. & q \ge 2. \end{array} \right. \label{eq:cubicspline}
\end{equation}
with normalisation $\sigma = [2/3,10/(7\pi),1/\pi]$.
This kernel satisfies the basic requirements
(\ref{eq:wtodelta}) and (\ref{eq:kernelnorm}), is even, has continuous first
derivatives and compact support of size $2h$. Smoother kernels can be introduced
by increasing the size of the compact support region (which correspondingly
increases the cost of evaluation by increasing the number of contributing
neighbours) and by using higher order interpolating spline functions. To this end the
quartic spline kernel
\begin{equation}
f(q) = \sigma \left\{ \begin{array}{ll}
(2.5-q)^4 - 5(1.5-q)^4 + 10(0.5-q)^4, & 0 \le q < 0.5; \\
(2.5-q)^4 - 5(1.5-q)^4, & 0.5 \le q < 1.5; \\
(2.5-q)^4, & 1.5 \le q < 2.5; \\
0. & q \ge 2.5. \end{array} \right. \label{eq:quarticspline} 
\end{equation}
 with normalisation $\sigma = [1/24,96/1199\pi,1/20\pi]$ and quintic spline kernel
\begin{equation}
f(q) = \sigma \left\{ \begin{array}{ll}
(3-q)^5 - 6(2-q)^5 + 15(1-q)^5, & 0 \le q < 1; \\
(3-q)^5 - 6(2-q)^5, & 1 \le q < 2; \\
(3-q)^5, & 2 \le q < 3; \\
0. & q \ge 3. \end{array} \right. \label{eq:quinticspline}
\end{equation}
 with normalisation $\sigma = [1/120,7/478\pi,1/120\pi]$ can be used \citep[e.g.][]{morrisphd}. The higher order polynomials have the advantage of
smoother derivatives which, in combination with the increased size of compact
support, decreases the sensitivity of the kernel to disorder in the particle
distribution (\S\ref{sec:kernelstability}).

\begin{figure}[ht!]
\begin{center}
\begin{turn}{270}\epsfig{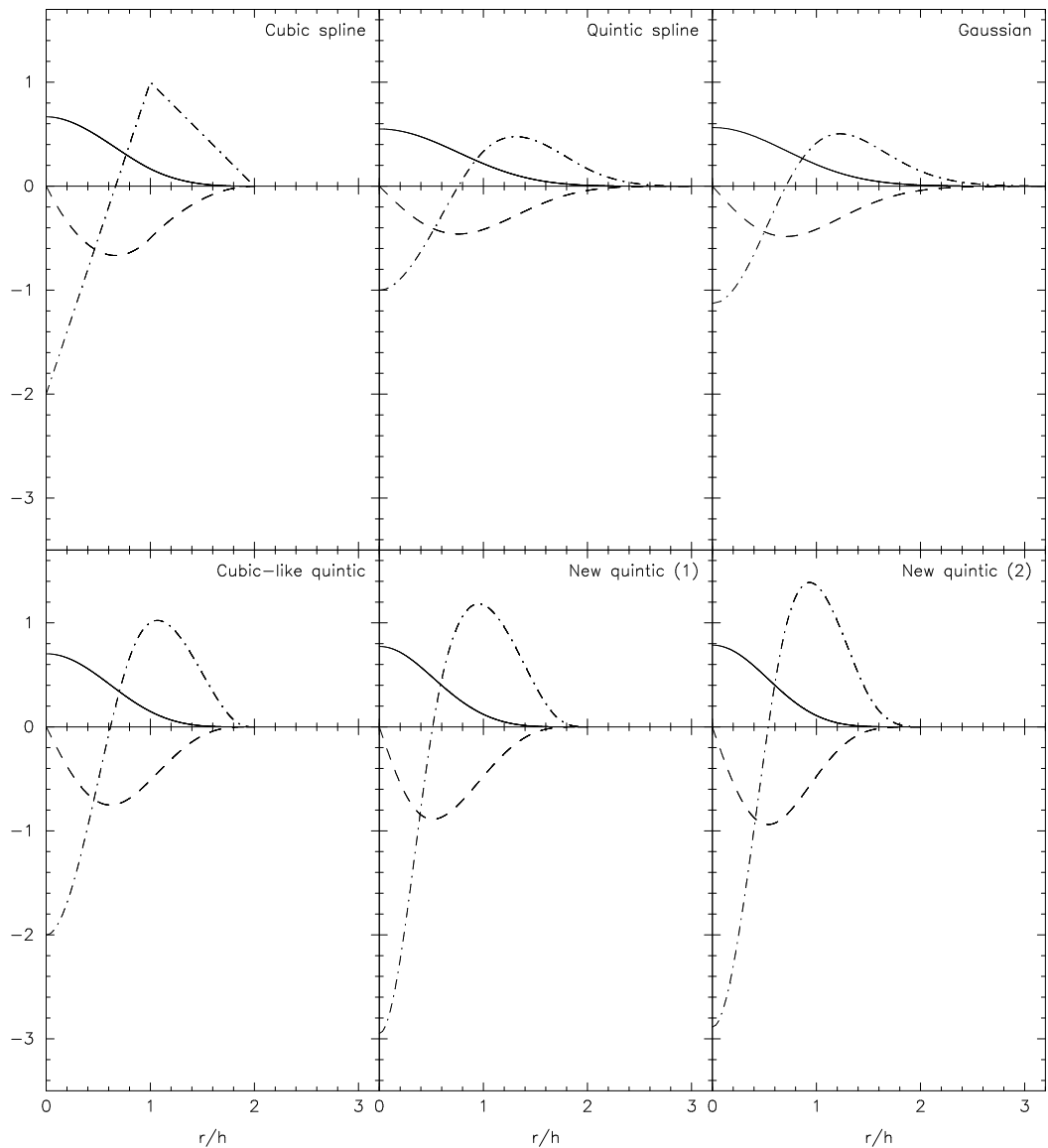}\end{turn}
\caption{Examples of SPH smoothing kernels (solid
line) together with their first (dashed) and second (dot-dashed) derivatives. Kernels
correspond to those given in the text. The cubic spline (top left) is the usual
choice, whilst the quintic (top, middle) represents a closer approximation to the Gaussian
kernel (top right), at the cost of increased compact support. The bottom row correspond to various quintic kernels with compact support
of $2h$ which we derive in
\S\ref{sec:DIYkernels}. The stability properties of all these kernels are compared
in Figure \ref{fig:kernelstability}.}
\label{fig:kernels}
\end{center}
\end{figure}

Note that it is entirely possible to construct kernels based on smoother
splines but which retain compact support of size $2h$. We derive a class of such
kernels and compare their stability properties with the kernels given in this
section in \S\ref{sec:DIYkernels}. In principle it is also possible to construct higher order kernels where the
second error term in (\ref{eq:kernelerrors}) is also zero. \citet{monaghan92} demonstrates that such higher order kernels may be
constructed from any lower order kernel such as (\ref{eq:cubicspline}) by the simple
relation
\begin{equation}
W_{high order} = B(1-Aq^2) W(q)
\end{equation}
where the parameters $A$ and $B$ are chosen to cancel the second moment and to
satisfy the normalisation condition (\ref{eq:kernelnorm}). The disadvantage of all such
kernels is that the kernel becomes negative in part of the domain which could
result in a negative density evaluation. Also it is not clear that such kernels
actually lead to significant improvements in accuracy in practical situations
(since the kernel is sampled at only a few points).

 From time to time various alternatives have been proposed to the kernel interpolation
at the heart of SPH, such as the use of Delaunay triangulations
\citep{pelupessyetal03} and normalisations of the kernel
interpolant (involving matrix inversion) which guarantee exact interpolations to arbitrary polynomial orders
\citep{mh03,bl99}. It remains to be seen whether any
such alternative proposals are viable in terms of the gain in accuracy versus
the inevitable increase in computational expense and algorithmic complexity.

 Finally we note that in most SPH codes, the kernel is evaluated by linear interpolation from a
pre-computed table of values, since kernel evaluations are computed
frequently. The computational cost involved in
calculating the kernel function is therefore the same whatever the functional
form. In the calculations given in this thesis, the kernel is tabulated as
$W(q)$ and $\partial W/\partial q$, where the
table is evenly spaced in $q^2$ to give a better interpolation in the outer
edges.

\subsection{A general class of kernels} \label{sec:DIYkernels}
 In this section we consider the possibility of constructing kernels based on
smoother splines than the cubic but which retain compact support of size $2h$. A general class of such
kernels may be derived by considering kernels of the form
\begin{equation}
f(q) = \sigma \left\{ \begin{array}{ll}
(r-q)^n + A(\alpha-q)^n + B(\beta-q)^n , & 0 \le q < \beta; \\
(r-q)^n + A(\alpha-q)^n, & \beta \le q < \alpha; \\
(r-q)^n, & \alpha \le q < r; \\
0. & q \ge r \end{array} \right. \label{eq:myquinticspline}
\end{equation}
where $n$ is the order, $r$ is the compact support size (in this case $r=2$), $A$ and $B$ are parameters to be
determined and $\alpha$ and $\beta$ are the two matching points (with $0 < \beta
< \alpha < r$), although an
arbitrary number of matching points could be added. The formulation
given above guarantees that the kernel and its derivatives are continuous at the
matching points and zero at the compact support radius $W(r)=dW/dq(r)=0$. To 
determine the parameters $A$ and $B$ we require two further constraints on the
form of the kernel. For the kernels to resemble the Gaussian, we constrain the kernel
gradient to be zero at the origin and also that the second derivative be minimum
at the origin (this also constrains $n\ge 3$), ie.
\begin{equation}
W'(0) = 0, \hspace{1cm} W'''(0) = 0. \label{eq:DIYkernelconstraints}
\end{equation}
 For the moment we leave the matching points as free parameters. From the conditions (\ref{eq:DIYkernelconstraints}), the
parameters $A$ and $B$ are given in terms of the matching points
by
\begin{equation}
A = \frac{r^{n-3}(r^2-\beta^2)}{\alpha^{n-3} (\alpha^2 - \beta^2)},
\hspace{1cm} B =  -\frac{r^{n-1} + A\alpha^{n-1}}{\beta^{n-1}}.
\end{equation}
In one dimension the normalisation constant is given by
\begin{equation}
\sigma = \frac{n+1}{2(A \alpha^{n+1} + B \beta^{n+1} + r^{n+1})}.
\end{equation}

As an example we can construct a quintic ($n=5$) kernel that
closely resembles the cubic spline kernel (\ref{eq:cubicspline}) in all but the continuity of the
second derivative. An example of such a kernel is given by the choice $\beta = 0.85$, $\alpha = 1.87$. This was chosen by
constraining the second derivative to be equal to that of the
cubic spline at the origin (ie. $W''(0) = -2$) and the turning point in the second derivative to be
located as close as possible to the that of the cubic spline ($W'''(q \approx 1) =
0$; note that an exact match is not possible under the constraints given).
This kernel is shown in Figure \ref{fig:kernels} (`cubic-like quintic'). The stability properties
are discussed in \S\ref{sec:kernelstability}.

However, it would be more interesting to investigate whether other kernels with
even better stability properties can be constructed. To this end we have
performed a survey of parameter space for quintic ($n=5$) kernels, from which we find that the most stable kernels are
those with matching points in the range $\beta \approx 0.5$ with $\alpha
\approx 1.7$ or $\beta \approx 0.7$ with $\alpha \approx 1.5$. These two kernels (`New Quintic(1)' and
`New Quintic (2)') are shown in Figure \ref{fig:kernels}. The
stability properties are discussed below.

\begin{figure}[t]
\begin{center}
\begin{turn}{270}\epsfig{file=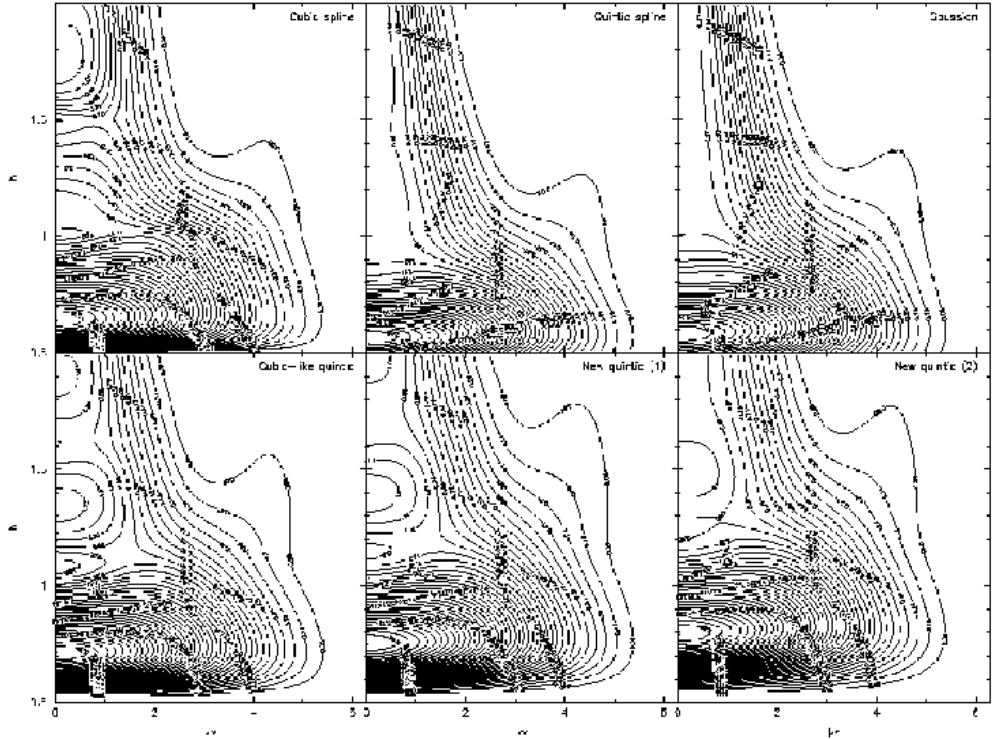,height=0.9\textwidth}\end{turn}
\caption{One dimensional stability properties of the kernels shown in Figure
\ref{fig:kernels} for isothermal SPH. The
$y$-axis gives the smoothing length in units of the particle spacing $\Delta x$, whilst the $x$-axis
corresponds to wavenumber in units of 1/$\Delta x$ (such that $kx \to 0$ represents the
limit of an infinite number of particles per wavelength and $h \to \infty$
represents the limit of an infinite number of neighbours). Contours
show the (normalised) square of the numerical sound speed from the dispersion relation
(\ref{eq:sphdispersion}). The quintic spline (top, centre) and Gaussian kernels show improved accuracy over the standard cubic spline kernel although at
a higher computational cost. The kernels derived in \S\ref{sec:DIYkernels} (bottom row) appear to give an
improvement in accuracy for $h \gtrsim 1.1$ although degrade rapidly for $h \lesssim 1.1$ where the
cubic spline retains a reasonable accuracy}
\label{fig:kernelstability}
\end{center}
\end{figure}

\subsection{Kernel stability properties}\label{sec:kernelstability}
 The accuracy of the kernels given in \S\ref{sec:kernels} and \S\ref{sec:DIYkernels} may
be compared via a stability analysis of the SPH equations. Detailed investigations of the stability properties of SPH have been given
elsewhere (e.g. \citealt{morrisphd}) and for this reason we refer the details of the stability
analysis to appendix \ref{sec:sphstability} (although as for the fluid equations, the
linearised form of the SPH equations are derived from a variational principle). The result for one-dimensional
SPH (for any equation of state) is the dispersion relation
\begin{eqnarray}
\omega^2_a & = & \frac{2m P_0}{\rho_0^2} \sum_b [1 - \cos{k (x_a - x_b)}]\pder{^2
W}{x^2}(x_a - x_b, h) \nonumber \\
& & + \frac{m^2}{\rho_0^2}\left(c_s^2 -  \frac{2P_0}{\rho_0}\right) \left[ \sum_b
\sin{k (x_a - x_b)}\pder{W}{x}(x_a - x_b, h) \right]^2,
\label{eq:sphdispersion}
\end{eqnarray}
where $c_s = \partial P / \partial \rho$ is the sound speed.
Figure \ref{fig:kernelstability} shows contours of the (normalised) square of the numerical sound speed
$C^2_{num} = \omega^2 / k^2$ as a function of wavenumber and smoothing length
(both in units of the average particle spacing). The sums in (\ref{eq:sphdispersion}) are calculated
numerically assuming an (isothermal) sound speed and particle spacing of unity (both wavelength and smoothing
length are calculated in units of the particle spacing). The quintic spline (top, centre) and the Gaussian (top right) show increasingly better stability
properties over the standard cubic spline (top left) although at increased computational expense.

 The stability properties of the `cubic-like' quintic kernel derived in \S\ref{sec:DIYkernels}
 (bottom left) are very similar to that of the cubic spline, except that the `trough' in the
contours of $C^2_{num}$ observed at $h=1.5\Delta p$ (where the closest neighbour crosses
the discontinuity in the second derivative) is much smoother. However, the accuracy
of this kernel appears to degrade for small smoothing lengths ($h \lesssim 1.1\Delta p$)
where the cubic spline retains a reasonable accuracy. Of the remaining two kernels derived in
\S\ref{sec:DIYkernels} (bottom centre and bottom right), the second
example (`New Quintic (2)') in particular appears to give slightly better accuracy than
the cubic spline over the range $h \gtrsim 1.1\Delta p$ although both kernels show the rapid
decline in accuracy for small smoothing lengths ($h \lesssim 1.1\Delta p$) observed in the
cubic-like quintic. It is worth noting that most multidimensional calculations use smoothing
lengths in the range $h = 1.1-1.2 \Delta p$. 

 In summary the new kernels appear to give a small gain in accuracy over the cubic spline
kernel, provided $h \gtrsim 1.1\Delta p$. However, the gain in accuracy from the use of these
alternative kernels is very minor compared to the substantial improvements in accuracy
gained by the incorporation of the variable smoothing length terms (\S\ref{sec:gradh}),
which effectively act as a normalisation of the kernel gradient.

\section{Fluid Equations}
\label{sec:fluideqs}

\subsection{Continuity equation}
\label{sec:sphcty}
The summation interpolant (\ref{eq:sumint}) takes a particularly simple form for the evaluation
of density, ie.
\begin{equation}
\rho_{a} = \smb W_{ab}.
\label{eq:rhosum}
\end{equation}
Taking the (Lagrangian) time derivative, we obtain
\begin{equation}
\frac{d\rho_a}{dt} = \smb (\bv_a -\bv_b) \cdot\gawab,
\label{eq:sphcty}
\end{equation}
which may be translated back to continuum form via the summation interpolant
(\ref{eq:sumint}) to give
\begin{eqnarray}
\frac{d\rho}{dt} & = & \bv\cdot\nabla \rho - \nabla\cdot(\rho \bv), \nonumber \\
& = & -\rho (\nabla\cdot\bv) \label{eq:ctycty}.
\end{eqnarray}
 This reveals that (\ref{eq:sphcty}) and therefore (\ref{eq:rhosum}) are SPH
expressions for the continuity equation. It is a remarkable fact that the entire
SPH formalism can be self-consistently derived using only (\ref{eq:rhosum}) in
conjunction with the first law of thermodynamics via a Lagrangian variational
principle. Such a derivation demonstrates that SPH has a robust Hamiltonian structure 
and ensures that the discrete equations reflect the symmetries inherent in the
Lagrangian, leading to the exact conservation of momentum, angular momentum and energy.

\subsection{Equations of motion}
\label{sec:sphmom}
 The Lagrangian for
Hydrodynamics is given by \citep{eckart60,salmon88,morrison98}
\begin{equation}
L = \int \left(\frac{1}{2}\rho \rv^2 - \rho u \right)\mathrm{dV},
\end{equation}
 where $u$ is the internal energy per unit mass. In SPH form this becomes
\begin{equation}
L = \sum_b m_b \left[\frac{1}{2} \rv_b^2 - u_b(\rho_b, s_b) \right],
\label{eq:Lsph}
\end{equation}
 where as previously we have replaced the volume element $\rho dV$ with the mass
per SPH particle $m$. We regard the particle co-ordinates as the canonical
variables. Being able to specify all of the terms in the Lagrangian directly in terms of
these variables means that the conservation laws will be automatically
satisfied, since the equations of motion then result from the
Euler-Lagrange equations
\begin{equation}
\frac{d}{dt}\left(\pder{L}{\bv_a}\right) - \pder{L}{\br_a} = 0.
\label{eq:el}
\end{equation}
 The internal energy is regarded as a function of the
particle's density, which in turn is specified as a function of the co-ordinates
by (\ref{eq:rhosum}). The terms in (\ref{eq:el}) are therefore given by
\begin{eqnarray}
\pder{L}{\bv_a} & = & m_a \bv_a, \label{eq:dLdva} \\
\pder{L}{\br_a} & = & \smb \left.\pder{u_b}{\rho_b}\right\vert_s \pder{\rho_b}{\br_a}.
\end{eqnarray}
From the first law of thermodynamics in the absence of dissipation we have
\begin{equation}
\left.\pder{u_b}{\rho_b}\right\vert_s = \frac{P_b}{\rho_b^2},
\label{eq:firstlawthermo}
\end{equation}
and using (\ref{eq:rhosum}) we have
\begin{equation}
\pder{\rho_b}{\br_a} = \smc \nabla_a W_{bc} \left(\delta_{ba} -
\delta_{ca}\right),
\end{equation}
such that
\begin{eqnarray}
\pder{L}{\br_a} & = & \smb \frac{P_b}{\rho_b^2} \smc \nabla_a W_{bc} \left(\delta_{ba} -
\delta_{ca}\right), \\
& = & m_a \smb \left(\frac{P_a}{\rho_a^2} + \frac{P_b}{\rho_b^2} \right)
\gawab,
\end{eqnarray}
where we have used the fact that the gradient of the kernel is anti-symmetric
(ie. $\nabla_a W_{ac} = -\nabla_a W_{ca}$). The SPH equation of motion in the
absence of dissipation is therefore given by
\begin{equation}
\frac{d\va}{dt} = - \smb \left( \frac{P_a}{\rho_a^2} +
\frac{P_b}{\rho_b^2}\right)\gawab,
\label{eq:sphmom}
\end{equation}
 which can be seen to explicitly conserve momentum since the contribution of the
summation to the momentum of particle $a$ is equal and opposite to that given to
particle $b$ (given the antisymmetry of the kernel gradient). Taking the time
derivative of the total angular momentum, we have
\begin{eqnarray}
\frac{d}{dt} \sum_a \br_a \times m_a \bv_a & = & \sma \left( \br_a \times \frac{d\bv_a}{dt}
\right), \\
& = & \sum_a \sum_b m_a m_b \left( \frac{P_a}{\rho_a^2} +
\frac{P_b}{\rho_b^2}\right) \br_a \times (\br_a - \br_b) F_{ab}, \nonumber \\
& = & -\sum_a \sum_b m_a m_b \left( \frac{P_a}{\rho_a^2} +
\frac{P_b}{\rho_b^2}\right) \br_a \times \br_b F_{ab}.
\end{eqnarray}
where the kernel gradient has been written as $\gawab = \br_{ab} F_{ab}$
This last expression is zero since the double summation is antisymmetric in $a$ and $b$ (this can be seen
by swapping the summation indices $a$ and $b$ in the double sum and adding half of
this expression to half of the original expression, giving zero). Angular momentum is
therefore also explicitly conserved.

\subsection{Energy equation}
\label{sec:sphenergy}
The energy equation also follows naturally from the
variational approach, where we may choose to integrate either the particle's internal
energy $u$, its specific energy $e$ or even its specific entropy $s$. Integrating the
specific energy guarantees that the total energy is exactly conserved and it is common practice to
use this quantity in finite difference schemes. However the usual argument against
this (which applies equally to finite difference schemes) is that in some circumstances (where the kinetic energy is much greater than
the thermal energy) the thermal energy can become negative by round-off error.
Integration of the specific entropy has some advantages and has been argued for in both SPH
\citep{sh02} and finite difference schemes (e.g. \citealt{bs99}).

\subsubsection{Internal energy}
The internal energy
equation in the absence of dissipation follows from the use of the first law of thermodynamics
(\ref{eq:firstlawthermo}), giving
\begin{equation}
\frac{du_a}{dt} = \frac{P_a}{\rho_a^2}\frac{d\rho_a}{dt}.
\end{equation}
Using (\ref{eq:sphcty}) therefore gives
\begin{equation}
\frac{du_a}{dt} = \frac{P_a}{\rho_a^2} \smb \bv_{ab}\cdot\gawab.
\label{eq:sphutherm}
\end{equation}

\subsubsection{Total energy}
The conserved (total) energy is found from the Lagrangian via the Hamiltonian
\begin{equation}
H = \sum_a \bv_a \cdot \pder{L}{\bv_a} - L,
\label{eq:H}
\end{equation}
where using (\ref{eq:dLdva}) and (\ref{eq:Lsph}) we have
\begin{equation}
H = \sma \left( \frac{1}{2}\rv_a^2 + u_a \right),
\label{eq:Hsph}
\end{equation}
which is simply the total energy of the SPH particles $E$ since the Lagrangian does
not explicitly depend on the time. Taking the (Lagrangian) time derivative of
(\ref{eq:Hsph}), we have
\begin{equation}
\frac{dE}{dt} = \sma \left(\bv_a \cdot \frac{d\bv_a}{dt} + \frac{du_a}{dt} \right).
\end{equation}
Substituting (\ref{eq:sphmom}) and (\ref{eq:sphutherm}) and rearranging we find 
\begin{equation}
\frac{dE}{dt} = \sma \frac{de_a}{dt} = \sum_a \sum_b m_a m_b \left( \frac{P_a}{\rho_a^2}\bv_b
+ \frac{P_b}{\rho_b^2}\bv_a \right)\cdot\gawab,
\end{equation}
and thus the specific energy equation (in the absence of dissipation) is given by
\begin{equation}
\frac{de_a}{dt} = \smb \left( \frac{P_a}{\rho_a^2}\bv_b
+ \frac{P_b}{\rho_b^2}\bv_a \right)\cdot\gawab.
\label{eq:sphenergy}
\end{equation}
Dissipative terms are discussed in \S\ref{sec:av}.

\subsubsection{Entropy}
 In the case of an ideal gas equation of state where
\begin{equation}
P = A(s) \rho^\gamma,
\end{equation}
the function $A(s)$ evolves according to
\begin{eqnarray}
\frac{dA}{dt} & = & \frac{\gamma-1}{\rho^{\gamma-1}}\left( \frac{du}{dt} -
\frac{P}{\rho^2}\frac{d\rho}{dt} \right), \nonumber \\
& = & \frac{\gamma-1}{\rho^{\gamma-1}}\left( \frac{du}{dt} \right)_{diss}.
\label{eq:sphentropy}
\end{eqnarray}
This has the advantage of placing strict controls on sources of entropy, since
$A$ is constant in the absence of dissipative terms. The thermal energy is
evaluated using
\begin{equation}
u = \frac{A}{\gamma - 1} \rho^{\gamma-1}.
\end{equation}
This formulation of the energy equation has been advocated in an SPH context by \citet{sh02}.

\subsection{Variable smoothing length terms}
\label{sec:gradh}
 The smoothing length $h$ determines the radius of interaction for each SPH
particle. Early SPH simulations used a fixed smoothing length for all particles.
However allowing each particle to have its own associated smoothing length which
varies according to local conditions increases the spatial resolution
substantially \citep{hk89,benz90}. The usual rule is to take
\begin{equation}
h_a \propto \left(\frac{1}{\rho_a} \right)^{(1/\nu)},
\label{eq:hrho}
\end{equation}
where $\nu$ is the number of spatial dimensions, although others are possible
\citep{monaghan00}. Implementing this rule
self-consistently is more complicated in SPH since the density $\rho_a$ is itself a
function of the smoothing length $h_a$ via the relation (\ref{eq:rhosum}). A simple
approach is to use the time derivative of (\ref{eq:hrho}),
\citep{benz90}, ie.
\begin{equation}
\frac{dh_a}{dt} = -\frac{h_a}{\nu\rho_a}\frac{d\rho}{dt},
\label{eq:hevol}
\end{equation}
which can then be evolved alongside the other particle quantities. This rule works well for most practical purposes, and maintains the relation (\ref{eq:hrho})
particularly well when the density is updated using the continuity equation
(\ref{eq:sphcty}). However, it has been known for
some time that, in order to be fully self-consistent, extra terms involving the
derivative of $h$ should be included in the momentum and energy equations (e.g.
\citealt{nelson94,np94,sea96}). Attempts to do this were, however, complicated to
implement \citep{np94} and therefore not generally adopted by the SPH
community. Recently \citet{sh02} have shown that the so-called $\nabla h$
terms can be self-consistently included in the equations of motion and energy
using a variational approach. \citet{sh02} included the variation of the
smoothing length in their variational principle by use of Lagrange multipliers,
however, in the context of the discussion given in \S\ref{sec:sphmom} we note
that by expressing the smoothing length as a function of $\rho$ we can therefore
specify $h$ as a function of the particle co-ordinates \citep{monaghan02}. That
is we have $h = h(\rho)$ where $\rho$ is given by
\begin{equation}
\rho_a = \smb W(\br_{ab}, h_a).
\label{eq:rhosumgradh}
\end{equation}
Taking the time derivative, we obtain
\begin{equation}
\frac{d\rho_a}{dt} = \frac{1}{\Omega_a} \smb \bv_{ab}\cdot\nabla_a W_{ab}(h_a),
\label{eq:sphctygradh}
\end{equation}
where
\begin{equation}
\Omega_a = \left[1 - \pder{h_a}{\rho_a}\smc
\pder{W_{ab}(h_a)}{h_a}\right].
\label{eq:omegaa}
\end{equation}
A simple evaluation of $\Omega$ for the kernel in the form
(\ref{eq:evenkernels}) shows that this term differs from unity even in the case
of an initially uniform density particle distribution (i.e. with constant
smoothing length). The effects of this correction term even in this simple case are investigated in the sound wave tests described in
\S\ref{sec:swavetests}.

The equations of motion in the hydrodynamic case may then be found using the Euler-Lagrange equations (\ref{eq:el})
and will therefore automatically conserve linear and angular momentum. The
resulting equations are given by \citep{sh02,monaghan02}
\begin{equation}
\frac{d\bv_a}{dt} = - \smb \left[ \frac{P_a}{\Omega_a\rho_a^2}\nabla_a
W_{ab}(h_a) + \frac{P_b}{\Omega_b\rho_b^2}\nabla_a W_{ab}(h_b)\right].
\label{eq:sphmomgradh}
\end{equation}

Calculation of the quantities $\Omega$ involve a summation over the particles
and can be computed alongside the density summation (\ref{eq:rhosumgradh}). To
be fully self-consistent we solve (\ref{eq:rhosumgradh}) iteratively to
determine both $h$ and $\rho$ self-consistently. We do this as follows: Using the predicted smoothing length from
(\ref{eq:hevol}), the density is initially calculated by a summation over the
particles. A new value of smoothing length $h_{new}$ is then computed from this
density using (\ref{eq:hrho}). Convergence is determined according to the
criterion
\begin{equation}
\frac{\vert h_{new} - h\vert}{h} < 1.0 \times 10^{-2}.
\end{equation}
For particles which are not converged, the density of (only) those particles
are recalculated (using $h_{new}$). This process is then repeated until all
particles are converged. Note that a particle's smoothing length is only set
equal to $h_{new}$ if the density is to be recalculated (this is to ensure that
the same smoothing length that was used to calculate the density is used to
compute the terms in the other SPH equations).  Also, the density only needs to
be recalculated on those particles which have not converged, since each
particle's density is independent of the smoothing length of neighbouring
particles. This requires a small adjustment to the density calculation routine
(such that the density can be calculated only for a selected list of particles,
rather than for all), but is relatively simple to implement and means that the
additional computational cost involved is negligible (at least for the problems
considered in this thesis). Note that in principle the calculated gradient
terms (\ref{eq:omegaa}) may also be used to implement an iteration scheme such
as the Newton-Raphson method which converges faster than our simple fixed point
iteration.

 Where the variable smoothing length terms are not explicitly calculated, we use
a simple averaging of the kernels and kernel gradients to maintain the symmetry
in the momentum and energy equations \citep{hk89,monaghan92}, ie.
\begin{equation}
W_{ab} = \frac{1}{2} \left[W_{ab}(h_a) + W_{ab}(h_b)\right],
\end{equation}
and correspondingly
\begin{equation}
\nabla_a W_{ab} = \frac{1}{2} \left[\nabla_a W_{ab}(h_a) + \nabla_a W_{ab}(h_b)\right].
\end{equation}
 Many of the test problems in this thesis are performed using this simple
formulation. This is
in order to show (particularly in the MHD case) that satisfactory results on the test problems are not
dependent on the variable smoothing length formulation. In almost every case,
however, self-consistent implementation of the variable smoothing length terms
as described above leads to a substantial improvement in accuracy
(demonstrated, for example, in \S\ref{sec:hydrotests} and in the MHD case in
\S\ref{sec:1Dtests}). Perhaps the only disadvantage to the full implementation
of the variable smoothing length terms is that the iterations of $h$ with $\rho$
mean that small density fluctuations are resolved by the method rather than
being smoothed out, which may be disadvantageous under some circumstances (e.g.
where the fluctuations are unphysical). One
possible remedy for this might be to use a slightly different relationship between
$h$ and $\rho$ than is given by (\ref{eq:hrho}). 

\section{Alternative formulations of SPH}
\label{sec:altforms}

 In \S\ref{sec:fluideqs} the SPH equations of motion and energy were derived from a
variational principle using only the density summation (\ref{eq:rhosum}) and the first law
of thermodynamics (\ref{eq:firstlawthermo}), leading to the equations of motion in the form
(\ref{eq:sphmom}) and the energy equation (\ref{eq:sphutherm}) or (\ref{eq:sphenergy}).
However many alternative formulations of the SPH equations are possible and have been used
in various contexts. In this section we demonstrate how such alternative formulations may
also be derived self-consistently using a variational principle.
 
 For example, a general form of the momentum equation in SPH is given by
 \citep{monaghan92}
\begin{equation}
\frac{d\bv_a}{dt} = -\smb \left( \frac{P_a}{\rho^{\sigma}_a \rho^{2-\sigma}_b} +
\frac{P_b}{\rho^\sigma_b \rho^{2-\sigma}_a} \right) \gwab,
\label{eq:genmom}
\end{equation}
which is symmetric between particle pairs for all choices of the parameter $\sigma$ and
therefore explicitly conserves momentum. \citet{rt01} use this form of the
momentum equation with $\sigma=1$ in their SPH formalism, finding that it gives slightly better results for
problems involving large density contrasts (they also use a slightly different procedure for
evaluating the density). \citet{mw03}, for similar reasons, use this equation with $\sigma = 3/2$,
citing a reduction in the relative error in the force calculation on particle $a$ due to
the influence of particle $b$ which is desirable in the case of particles with large density
differences. However, it is apparent from the derivation given in \S\ref{sec:sphmom} that forms of
this equation other than the standard $\sigma = 2$ case cannot be derived
consistently using the density summation (\ref{eq:rhosum}) and correspondingly the continuity
equation in the form (\ref{eq:sphcty}). We are therefore led to the
conclusion that a consistent formulation of the SPH equations using the general
form of the momentum equation given above must involve modification of the
continuity equation in some way. We show below that the general form of the continuity equation which is consistent
with (\ref{eq:genmom}) is derived from the continuum equation
\begin{equation}
\frac{d\rho}{dt} = -\rho \nabla\cdot \bv,
\end{equation}
expressed in the form
\begin{equation}
\frac{d\rho}{dt} = \rho^{2-\sigma} \left[\bv\cdot\nabla (\rho^{\sigma-1}) -
\nabla\cdot (\bv \rho^{\sigma-1}) \right],
\end{equation}
with SPH equivalent
\begin{equation}
\frac{d\rho_a}{dt} = \rho^{2-\sigma}_a \sum_b m_b
\frac{(\bv_a - \bv_b)}{\rho^{2-\sigma}_b}\cdot \gwab.
\label{eq:gencty}
\end{equation}

 In order to demonstrate that this is so, we use this expression for the density to derive the equations of motion
and energy via a variational principle. 

\subsection{Variational principle}
In the derivation given in
\S\ref{sec:sphmom}, the variables in
the Lagrangian were explicitly written as a function of the particle co-ordinates (via the identity
\ref{eq:rhosum}), guaranteeing the exact conservation of linear and angular 
momentum in the equations of motion via the use of the Euler-Lagrange equations.
Using a more general form of the continuity equation, however, means that the
density can no longer be expressed directly as a function of the particle
co-ordinates and therefore that the derivation given in the previous section
cannot be applied in this case. However we may still use the Lagrangian
to derive the equations of motion by introducing
constraints on $\rho$ in a manner similar to that of \citet{bl99}. In this case
conservation of momentum and energy can be shown to depend on the formulation of the velocity
terms in the continuity equation (in particular that the term should be
expressed as a velocity difference). Clearly the major disadvantage of using a
continuity equation of any form rather than the SPH summation is that mass is no
longer conserved exactly. It is shown in \S\ref{sec:spmhdmom} that the kind of
variational principle given below may also be used to derive the equations of motion
and energy in the MHD case.

For stationary action we require 
\begin{equation}
\delta \int L \mathrm{dt} = \int \delta L \mathrm{dt} = 0,
\label{eq:varprin}
\end{equation}
where we consider variations with respect to a small change in the particle
co-ordinates $\delta \br_a$. We therefore have
\begin{equation}
\delta L = m_a \bv_a\cdot\delta\bv_a - \smb\left.\pder{u_b}{\rho_b}\right\vert_s \delta\rho_b.
\label{eq:deltaL}
\end{equation}
The Lagrangian variation in density is given, from (\ref{eq:gencty}), by 
\begin{equation}
\delta\rho_b  = \rho^{2-\sigma}_b \sum_c \frac{m_c}{\rho^{2-\sigma}_c} \left(\delta \br_b - \delta
\br_c\right)\cdot \nabla_b W_{bc}.
\label{eq:deltarho} 
\end{equation}

Using (\ref{eq:deltarho}) and the first law of thermodynamics (\ref{eq:firstlawthermo}) in (\ref{eq:deltaL}) and rearranging, we find
\begin{equation}
\frac{\delta L}{\delta \br_a} = -\smb \frac{P_b}{\rho_b^\sigma}\sum_c\frac{m_c}{\rho^{2-\sigma}_c}\nabla_b W_{bc} (\delta_{ba} -
	 \delta_{ca}).
\end{equation}
Putting this back into
(\ref{eq:varprin}), integrating the velocity term by
parts and simplifying (using $\gwab = -\nabla_b W_{ba}$), we obtain
\begin{equation}
\int\bigg{[}  - m_a \frac{d\bv_a}{dt} - \smb \left(\frac{P_a}{\rho_a^\sigma
\rho_b^{2-\sigma}} + \frac{P_b}{\rho_b^\sigma\rho_a^{2-\sigma}} \right) \nabla_a W_{ab} 
\bigg{]}\delta\br_a \mathrm{dt} = 0, 
\end{equation}
from which we obtain the momentum equation in the form (\ref{eq:genmom}).
This equation is therefore consistent with the continuity equation in
the form (\ref{eq:gencty}). In the particular case considered by \citet{mw03} 
($\sigma=3/2$) this would imply a discrete form of the continuity equation given
by
\begin{equation}
\frac{d\rho_a}{dt} = \sqrt{\rho_a} \smb \frac{\bv_{ab}}{\sqrt{\rho_b}} \cdot \gwab.
\end{equation}
\citet{mw03} choose to retain the use of the usual SPH summation (\ref{eq:rhosum}) to determine the
density. In the case considered by \citet{rt01} ($\sigma=1$), the
continuity equation becomes
\begin{equation}
\frac{d\rho_a}{dt} = \rho_a \smb \frac{\bv_{ab}}{\rho_b} \cdot \gwab,
\label{eq:altcty}
\end{equation}  
which is again somewhat different to the density estimation used in their paper. The
continuity equation (\ref{eq:altcty}), when used in conjunction with the
appropriate formulation of the momentum equation, has some
advantages in the case of fluids with large density differences (e.g. at a
water/air interface) since the term inside the summation involves only the particle
volumes $m/\rho$ rather than their mass, with the effect that large mass
differences between individual particles have less influence on the calculation of the
velocity divergence (Monaghan, private communication). An alternative is the formalism
proposed by \citet{os03}, which we discuss in \S\ref{sec:ottschnetter}.

 The internal energy equation consistent with the general momentum
equation (\ref{eq:genmom}) is given by
\begin{equation}
\frac{du_a}{dt} = \frac{P_a}{\rho_a^{\sigma}} \smb \frac{\bv_{ab}}{\rho_b^{2-\sigma}} \cdot\gwab,
\end{equation}
which is indeed the formalism used by \citet{mw03} (with $\sigma = 3/2$) since it was found,
unsurprisingly in this context, that
integration of this equation resulted in much less numerical noise than using
other formalisms of the internal energy equation (in conjunction with their use
of (\ref{eq:genmom}) with $\sigma=3/2$ as the momentum equation). The
form of the total energy equation consistent with (\ref{eq:genmom}) and
(\ref{eq:gencty}) is given by
\begin{equation}
\frac{de_a}{dt} = -\smb
\left(\frac{P_a}{\rho_a^\sigma\rho_b^{2-\sigma}} \bv_b
+ \frac{P_b}{\rho_b^\sigma \rho_a^{2-\sigma}} \bv_a \right) \cdot \nabla_a W_{ab}.
\end{equation}
We note the energy equation used by \citet{rt01} is different to the formulation
given above (with $\sigma=1$) and therefore variationally inconsistent with their implementation of the
momentum equation. \citet{hk89}
point out that inconsistencies between the forms of the energy and momentum equations
result in errors of $\mathcal{O}(h^2)$ in the energy conservation. In this sense
the difference between a consistent and inconsistent formalism is fairly minor,
although a consistent formulation between the momentum and energy equations
in general appears to lead to slightly improved results (as found by
\citeauthor{mw03}). In practise we find that using alternative formulations of the continuity
equation generally gives slightly worse results than (even inconsistent) use of the density
summation.

\subsection{General alternative formulation}
 The momentum equation (\ref{eq:genmom}) can be generalised still further
by noting that the continuity equation (\ref{eq:ctycty}) can be written as
\begin{equation}
\frac{d\rho}{dt} = \phi \left[ \bv\cdot\nabla \left(\frac{\rho}{\phi}\right) - \nabla\cdot
\left( \frac{\rho \bv}{\phi} \right) \right],
\end{equation}
with SPH equivalent
\begin{equation}
\frac{d\rho_a}{dt} = \phi_a \smb \frac{\bv_{ab}}{\phi_b} \cdot \gwab,
\label{eq:vgencty}
\end{equation}
where $\phi$ is \emph{any} scalar variable defined on the particles. Deriving the
momentum equation consistent with this equation in the manner given above we find
\begin{equation}
\frac{d\bv_a}{dt} = - \smb \left( \frac{P_a}{\rho_a^2}\frac{\phi_a}{\phi_b} +
\frac{P_b}{\rho_b^2}\frac{\phi_b}{\phi_a}\right)\gwab,
\label{eq:vgenmom}
\end{equation}
which conserves momentum for any choice of $\phi$. In the case given in the
previous section we would have $\phi = \rho^{2-\sigma}$. Choosing $\phi = \rho/\sqrt{P}$ gives
\begin{equation}
\frac{d\bv_a}{dt} = - \smb \left( 2\frac{\sqrt{P_a P_b}}{\rho_a \rho_b} \right)\gwab.
\label{eq:hkmom}
\end{equation}
which is the momentum equation used by \citet{hk89}. The continuity equation
consistent with this form is therefore
\begin{equation}
\frac{d\rho_a}{dt} = \frac{\rho_a}{\sqrt{P_a}} \smb \frac{\sqrt{P_b}}{\rho_b}
\bv_{ab}\cdot\gwab,
\end{equation}
which at first sight appears somewhat bizarre, although it is certainly a valid
expression of the continuity equation in SPH form. It is unclear whether using
such alternative formulations of the continuity equation, in the name of
consistency, has any advantages over the usual density summation. 
We leave it as an exercise for the reader to amuse themselves by exploring various
other combinations of variables, noting that the forms of the internal and
total energy equations consistent with (\ref{eq:vgencty}) and (\ref{eq:vgenmom}) are given by
\begin{equation}
\frac{du_a}{dt} = \frac{P_a}{\rho_a^2} \smb \frac{\phi_a}{\phi_b} \bv_{ab} \cdot
\gwab,
\end{equation}
and
\begin{equation}
\frac{de_a}{dt} = - \smb \left( 
\frac{P_a}{\rho_a^2}\frac{\phi_a}{\phi_b} \bv_b +
\frac{P_b}{\rho_b^2}\frac{\phi_b}{\phi_a} \bv_a\right) \cdot \gwab.
\end{equation}

\subsection{\citeauthor{os03} formulation}
\label{sec:ottschnetter}
 Other formulations of the SPH equations have also been proposed to deal with
the problem of large density gradients.
For example \citet{os03} propose modifying the SPH summation to give
\begin{eqnarray}
n_a & = & \sum_b W_{ab}, \nonumber \\ 
\rho_a & = & m_a n_a,
\label{eq:ossum}
\end{eqnarray}
that is where the number density of particles $n$ is calculated by summation rather
than the mass density $\rho$. This is to improve the interpolation when particles of
large mass differences interact. Taking the time derivative of (\ref{eq:ossum}),
the continuity equation is given by (as in \citealt{os03})
\begin{equation}
\frac{d\rho_a}{dt} = m_a\sum_b \bv_{ab}\cdot\gwab.
\label{eq:oscty}
\end{equation}
 For equal mass particles this formalism is
exactly the same as the usual summation (\ref{eq:rhosum}).  The formulation (\ref{eq:ossum}) enables the density to be expressed as a
function of the particle co-ordinates and thus the derivation of the equations of
motion and energy can be done in a straightforward manner using the Euler-Lagrange equations, as in
\S\ref{sec:sphmom}. The resulting equation of motion is given by
\begin{equation}
m_a \frac{d\bv_a}{dt} = -\sum_b \left( \frac{P_a}{n_a^2} + \frac{P_b}{n_b^2}
\right) \gwab,
\end{equation}
which is somewhat different to the equation of motion used in \citet{os03}
 (they use the form \ref{eq:genmom} with $\sigma=1$). The internal
energy equation follows from the continuity equation (\ref{eq:oscty}) and the first law of
thermodynamics (\ref{eq:firstlawthermo}). We find
\begin{equation}
m_a \frac{du_a}{dt} = \frac{P_a}{n^2_a} \sum_b \bv_{ab}\cdot\gwab.
\end{equation}
\citet{os03} use a formulation of the internal energy
equation where the pressure term is symmetrised, which is inconsistent
with their use of (\ref{eq:ossum}). 
The total energy equation consistent with their formalism can also be derived using
the Hamiltonian (\S\ref{sec:sphenergy}) and is given by
\begin{equation}
m_a \frac{de_a}{dt} = -\sum_b \left( \frac{P_a}{n_a^2}\bv_b +
\frac{P_b}{n_b^2}\bv_a
\right) \cdot \gwab.
\end{equation}
In this case use of the self-consistent formalism presented above should lead to
slightly improved results over the momentum and energy equations employed by
\citet{os03}, since the density is still calculated via a direct summation over
the particles.

\section{Shocks}
\label{sec:av}
 In any high-order numerical scheme, the simulation of shocks is
accompanied by unphysical oscillations behind the shock front. This occurs
because in discretising the continuum equations (in the SPH case using
\ref{eq:sumint}) we assume that the fluid quantities are smoothly varying on the
smallest length scale (in SPH this is the smoothing length $h$). This means that
discontinuities on such scales are not resolved by the numerical method. 
The simplest approach to this problem is to introduce a
small amount of viscosity into the simulation which
acts to spread out the shock front so that it can be sufficiently resolved \citep{vr50,rm67}. This is similar to the way in which shock
fronts are smoothed out by nature, although in the latter case the effect occurs
at a much finer level.  The disadvantage of using such an `artificial' viscosity is
that it can produce excess heating elsewhere in the simulation.
As such the use of artificial viscosity is regarded by many numerical practitioners
as outdated since most finite difference schemes now rely on methods which
either restrict the magnitude of the numerical flux across a shock front in
order to prevent unphysical oscillations (such as total variation diminishing (TVD)
schemes) or by limiting the jump in the basic variables across the shock front using the
exact solution to the Riemann problem (Godunov-type schemes). There remain, however,
distinct advantages to the use of an
artificial viscosity, primarily that, unlike the Godunov-type schemes, it is easily applied where new physics is
introduced (such as a more complicated equation of state than the ideal gas
law) and the
complexity of the algorithm does not increase with the number of spatial dimensions.
In the case of magnetohydrodynamics, artificial viscosity is commonly used even in
standard finite-difference codes\footnote{for example in the widely used ZEUS code for
astrophysical fluid dynamics \citep{sn92}} since the Riemann problem is difficult to solve and
computationally expensive. Furthermore, dissipative terms are often still used
even when a Riemann solver has been implemented (e.g. \citealt{balsara98}).
For these reasons artificial viscosity
methods continue to find widespread usage, particularly in simulations using unstructured or Lagrangian
meshes \citep*{csw98}. 

In recent years it has been shown that Godunov-type schemes can in fact be used in conjunction 
with SPH by regarding interacting particle pairs as
left and right states of the Riemann problem \citep{cw03,inutsuka02,pm02,monaghan97}. In this manner the implementation of Godunov-type schemes to multidimensional
problems is greatly simplified in SPH because the one-dimensional Riemann
problem is solved between particle pairs, removing the
need for complicated operator splitting procedures in higher dimensions. The formalism
presented by \citet{cw03} is remarkably simple to incorporate into any standard
SPH code. A Godunov-type scheme for MHD in SPH would be extremely useful (although not
widely applicable), but it is well beyond the scope of this thesis. We therefore
formulate artificial dissipation terms using the formulation of \citet{monaghan97} which is
generalised to the MHD case in \S\ref{sec:mhdav}. The problem of excess heating is addressed
by the implementation of switches to turn off the dissipative terms away from shock
fronts, described in \S\ref{sec:avlim}.

\subsection{Artificial viscosity and thermal conductivity}
\label{sec:avformulation}
 A variety of different formulations of artificial viscosity in SPH have been
used, however the most common implementation is that given by \citet{monaghan92}, where the
term in equation (\ref{eq:sphmom}) is given by
\begin{equation}
\left(\frac{d\bv_a}{dt} \right)_{diss} = \smb \frac{-\alpha \bar{c}_{ab}\mu_{ab} +
\beta\mu^2_{ab}}{\bar{\rho}_{ab}} \gawab, \hspace{2cm} \mu_{ab} = \frac{h \bv_{ab}\cdot\br_{ab}}{\br_{ab}^2 + 0.01h^2},
\label{eq:avm92}
\end{equation}
where $\bv_{ab} \equiv \bv_a - \bv_b$ (similarly for $\br_{ab}$), barred quantities
refer to averages between particles $a$ and $b$, and $c$ refers to the sound speed.
This viscosity is applied only when the particles are in compression (ie.
$\bv_{ab}\cdot\br_{ab} < 0$), is Galilean invariant, conserves total linear and
angular momentum and vanishes for rigid body rotation. The $\beta$ term (quadratic
in $\bv_{ab}$) represents a form of
viscosity similar to the original formulation of \citet{vr50} and becomes dominant in the limit of large velocity differences (ie. in high
Mach number shocks). The $\alpha$ term is linear in
$\bv_{ab}$ and is dominant for small velocity differences\footnote{The
introduction of such a term into artificial viscosity methods is generally
attributed to \citet{landshoff55} (see, e.g. \citealt{csw98})}. Most astrophysical SPH implementations follow \citet{monaghan92}
in setting $\alpha = 1$ and $\beta = 2$ which provides the necessary dissipation near a shock front.

 The term given by equation (\ref{eq:avm92}) was constructed to
have the properties described above, however in the relativistic case it was
unclear as to what form such an artificial viscosity should take. \citet{cm97}
thus formulated an artificial viscosity for ultra-relativistic shocks in SPH by
analogy with Riemann solvers. This is outlined by \citet{monaghan97}
in a discussion of SPH and Riemann solvers. The essential idea is to regard the
interacting particles as left and right Riemann states and to construct a
dissipation which involves jumps in the physical variables. The dissipation term in the
force (giving artificial viscosity) therefore involves a jump in the velocity
variable and is similar to (\ref{eq:avm92}), taking the form (for $\bv_{ab}\cdot\br_{ab} < 0$)
\begin{equation}
\left(\frac{d\bv_a}{dt} \right)_{diss} = -\smb \frac{ \alpha\rv_{sig} (\bv_a -
\bv_b) \cdot\runit}{2 \bar{\rho}_{ab}} \gawab,
\label{eq:Piab}
\end{equation}
where $\rv_{sig}$ is a signal velocity and $\runit \equiv (\br_a-\br_b)/ \vert \br_a -
\br_b\vert $ is a unit vector along
the line joining the particles. Note that this formalism differs from
(\ref{eq:avm92}) in that a factor of $h/\vert \br_{ab} \vert$ has been removed. Also the $0.01h^2$
term has been removed from the denominator since for variable smoothing lengths
it is unnecessary. The jump in velocity involves only the component along the
line of sight since this is the only component expected to change at a shock front.
In a similar manner, the dissipative term in the specific energy
equation (\ref{eq:sphenergy}) is given by
\begin{equation}
\left( \frac{de_a}{dt} \right)_{diss} = - \smb \frac{\rv_{sig} (e^*_a
- e^*_b)}{2\bar{\rho}_{ab}} \runit \cdot \gawab,
\label{eq:dendtdiss}
\end{equation}
where $(e^*_a - e^*_b)$ is the jump in specific energy. The specific
energy used in this term is given by
\begin{equation}
e^*_a = \left \lbrace \begin{array}{ll} \frac{1}{2}\alpha(\bv_a\cdot\runit)^2 + \alpha_u u_a, & \bv_{ab}\cdot\br_{ab} < 0; \\
\alpha_u u_a & \bv_{ab}\cdot\br_{ab} \ge 0; \end{array} \right.
\label{eq:ediff}
\end{equation}
that is, where the specific kinetic energy has been
projected along the line joining the particles, since only the component of
velocity parallel to this vector is expected to jump at a shock front. Note that in general we use a
different parameter $\alpha_u$ to control the thermal energy term and that this term is applied to
particles in both compression and rarefaction.

 The signal velocity represents the maximum speed of signal propagation along
the line of sight between the two particles. Whilst many formulations could be
devised, it turns out that the results are not sensitive to the particular
choice made. A simple estimate of the signal velocity is given by
\begin{equation}
\rv_{sig} = c_a + c_b - \beta\bv_{ab}\cdot\runit
\end{equation}
where $c_a$ denotes the speed of sound of particle $a$ and $\beta \sim 1$, such
that $\rv_{sig}/2$ is an estimate of the maximum speed for linear wave propagation
between the particles. The $\beta$ term, which acts as a von Neumann and Richtmyer
viscosity as in equation (\ref{eq:avm92}), arises naturally in this formulation.
Practical experience suggests, however, that $\beta = 2$ is a better choice. For
a more general discussion of signal velocities we refer the reader to
\citet{monaghan97} and \citet{cm97}.

 The contribution to the thermal energy from the dissipative
terms is found using
\begin{equation}
\frac{du_a}{dt} = \frac{de_a}{dt} - \bv_a \cdot \frac{d\bv_a}{dt}.
\end{equation}
 In this case we obtain
\begin{equation}
\left(\frac{du_a}{dt} \right)_{diss} = \smb \frac{
\rv_{sig}}{2\bar{\rho}_{ab}} \left\{ -\frac{1}{2}\alpha\left[(\bv_a - \bv_b)\cdot\runit
\right]^2 + \alpha_u (u_a - u_b) \right\}
\runit \cdot \gawab
\label{eq:dudtdiss}
\end{equation}
which is added to the non-dissipative term (\ref{eq:sphutherm}). The first term
is the positive definite contribution to the thermal energy from the artificial
viscosity (since the kernel gradient is always
negative). The second term (involving a jump in thermal energy) provides an
artificial thermal conductivity. Physically this means that discontinuities in
the thermal energy are also smoothed.

 The artificial dissipation given by (\ref{eq:Piab})-(\ref{eq:dudtdiss}) is used as a basis for constructing an
appropriate dissipation for the MHD case in \S\ref{sec:mhdav}.

\subsection{Artificial dissipation switches}
\label{sec:avlim}

\subsubsection{Artificial viscosity}
 In both (\ref{eq:avm92}) and (\ref{eq:Piab}) the artificial viscosity is applied universally across the particles despite only
being needed when and where shocks actually occur. This results in SPH simulations
being much more dissipative than is necessary and can cause problematic effects 
where this dissipation is unwanted (such as in the presence of shear flows). 
 A switch to reduce the artificial viscosity away from shocks is given by
\citet{mm97}. Using this switch in multi-dimensional simulations substantially reduces the
problematic effects of using an artificial viscosity in SPH. 
 
 The key idea is to regard the dissipation parameter $\alpha$ (c.f. equation
\ref{eq:Piab}) as a particle property. This can then be evolved along with the
fluid equations according to
\begin{equation}
\frac{d\alpha_a}{dt} = -\frac{\alpha_a-\alpha_{min}}{\tau_a} + \mathcal{S}_a,
\label{eq:daldt}
\end{equation}
such that in the absence of sources $\mathcal{S}$, $\alpha$ decays to a value
$\alpha_{min}$ over a timescale $\tau$. The timescale $\tau$ is calculated
according to
\begin{equation}
\tau = \frac{h}{\mathcal{C}\rv_{sig}},
\end{equation}
where $h$ is the particle's smoothing length, $\rv_{sig}$ is the maximum signal
propagation speed at the
particle location and $\mathcal{C}$ is a dimensionless parameter with value $0.1
< \mathcal{C} <
0.2$. We conservatively use $\mathcal{C}=0.1$ which means that the value of $\alpha$ decays
to $\alpha_{min}$ over
$\sim 5$ smoothing lengths.

 The source term $\mathcal{S}$ is chosen such that the artificial dissipation grows
as the particle approaches a shock front. We use \citep{rosswogetal00}
\begin{equation}
\mathcal{S} = \mathrm{max}(-\nabla\cdot\bv, 0)(2.0 - \alpha),
\label{eq:alphasource}
\end{equation} 
such that the dissipation grows in regions of strong compression. Following \citet{mm97} where
the ratio of specific heats $\gamma$ differs from 5/3 (but not for the isothermal case), we multiply $\mathcal{S}$ by a factor
\begin{equation}
\left[\mathrm{ln}\left( \frac{5/3 +1}{5/3-1} \right)\right] /
\left[\mathrm{ln}\left( \frac{\gamma+1}{\gamma-1}\right)\right]
\end{equation}

 The source term is multiplied by a factor $(2.0 - \alpha)$ as the standard source term
given by \citet{mm97} was found to produce insufficient damping at shock fronts when used in
conjunction with the \citet{monaghan97} viscosity. The source term (\ref{eq:alphasource})
is found to provide sufficient damping on the \citet{sod78} hydrodynamic shock tube problem and in
the MHD shock tube tests we describe in chapter \S\ref{sec:1Dtests} (ie. $\alpha_{max} \sim 1$ for these problems).
In order to conserve momentum the average value $\bar{\alpha}=0.5(\alpha_a + \alpha_b)$
is used in equations (\ref{eq:Piab}), (\ref{eq:ediff}) and (\ref{eq:dudtdiss}). A lower limit of
$\alpha_{min}=0.1$ is used to preserve order away from shocks (note that this is an
order of magnitude reduction from the usual value of $\alpha=1.0$ everywhere).

 The numerical tests in \S\ref{sec:1Dtests} demonstrate that use of this switch gives a significant
reduction in dissipation away from shocks whilst preserving the
shock-capturing ability of the code.

\subsubsection{Artificial thermal conductivity}
\label{sec:condswitch}
 A similar switch to that used in the
artificial viscosity may therefore be devised for the artificial thermal
conductivity term, with the parameter $\alpha_u$ evolved according to
\begin{equation}
\frac{d\alpha_{u,a}}{dt} = -\frac{\alpha_{u,a} - \alpha_{u,min}}{\tau_a} + \mathcal{S}_a,
\label{eq:daludt}
\end{equation}
where the decay timescale $\tau$ is the same as that used in (\ref{eq:daldt})
and in this case we use $\alpha_{u,min} = 0$. The corresponding source term is given by
\begin{equation}
\mathcal{S} = \vert \nabla \sqrt{u} \vert, 
\end{equation}
which is constructed to have dimensions of inverse time. The gradient term is computed according to
\begin{equation}
\nabla \sqrt{u} = \frac12 u^{-1/2} \nabla u,
\end{equation}
where
\begin{equation}
\nabla u_a  = \frac{1}{\rho_a}\smb (u_a - u_b) \nabla_a W_{ab}(h_a).
\end{equation} 
 Use of this switch ensures that artificial thermal
conductivity is only applied at large gradients in the thermal energy.
The need to do so in dissipation-based shock capturing schemes is often
concealed by smoothing of the initial conditions in shock tube tests
(\S\ref{sec:sodshock}). From the first law of
thermodynamics (\ref{eq:firstlawthermo}) we infer that gradients in the 
thermal energy correspond to large gradients in the density. In a hydrodynamic shock
these occur either at the shock front or at the contact discontinuity. Artificial
viscosity is not required at the contact discontinuity because the pressure is constant
across it. Using unsmoothed initial conditions and in the absence of artificial thermal conductivity, a significant overshoot in
thermal energy occurs at the contact discontinuity (this phenomenon is known as `wall
heating' and is illustrated in Figure \ref{fig:sod_unsmoothed_av}). The resulting glitch in
the pressure is often ascribed to `starting errors' due to the unsmoothed
initial conditions. However, applying
smoothing to the initial conditions of a shock-tube test means that gradients across the 
contact discontinuity
remain smoothed throughout the evolution (see e.g. Figure \ref{fig:sod_smoothed_av}), removing the need for artificial thermal
conductivity which acts to spread gradients in the thermal energy. Whilst there is
also a gradient in thermal energy at a shock front, this is smoothed out
by the application of artificial viscosity there and so the need for artificial thermal
conductivity can go unnoticed. In \S\ref{sec:sodshock} we
present results of the standard \citet{sod78} shock tube test, showing the
effectiveness of the switch discussed above in applying the requisite amount of
smoothing at the contact discontinuity.

\section{Timestepping}
\label{sec:timestep}

\subsection{Predictor-corrector scheme}
We integrate the SPH equations in this thesis using a slight modification of the
standard predictor-corrector (Modified Euler) method which is second order
accuracy in time \citep{monaghan89}.
The predictor step is given by
\begin{eqnarray}
\bv^{1/2} & = & \bv^0 + \frac{\Delta t}{2} {\bf f}^{0}, \\
\br^{1/2} & = & \br^0 + \frac{\Delta t}{2} {\bf v}^{1/2}, \\
e^{1/2} & = & e^0 + \frac{\Delta t}{2} {\dot{e}}^{0},
\end{eqnarray}
where in practice we use ${\bf f}^{0} \approx {\bf f}^{-1/2}$ and ${\dot{e}}^{0}
\approx  {\dot{e}}^{-1/2}$ to give a one-step method. The rates of change of these quantities are then computed via the SPH
summations using the predicted values at the half step, ie.
\begin{equation}
{\bf f}^{1/2} = {\bf f}(\br^{1/2},\bv^{1/2}) \hspace{2cm} \dot{e}^{1/2} =
\dot{e}(\br^{1/2},\bv^{1/2})
\end{equation}
The corrector step is given by
\begin{eqnarray}
\bv^{*} & = & \bv^0 + \frac{\Delta t}{2} {\bf f}^{1/2}, \\
\br^{*} & = & \br^0 + \frac{\Delta t}{2} {\bf v}^{*}, \\
e^{*} & = & e^0 + \frac{\Delta t}{2} {\dot{e}}^{1/2},
\end{eqnarray}
and finally
\begin{eqnarray}
\bv^{1} & = & 2\bv^* - \bv^0,\\
\br^{1} & = & 2\br^* - \br^0, \\
e^{1} & = & 2e^* - e^0.
\end{eqnarray}
 Note that in this scheme the position updates in both the predictor and corrector steps use
the updated value of velocity. This effectively means that the position is
updated using both the first and second derivatives. From numerical experiments we find that this
scheme gives much better stability properties. Where evolved,  density,
smoothing length, magnetic field and the dissipation parameters follow the
energy evolution. The total energy $e$ is interchangeable for the thermal
energy $u$. 

\subsection{Reversible integrators}
 The simple predictor-corrector method given above is adequate for all the
problems considered in this thesis since the integration time is quite short. For
large simulations over long timescales, however, the accuracy and stability of the
integration method needs more careful attention. In the past decade or so a
substantial research effort has been devoted to the
development of high accuracy so-called `geometric' integrators for Hamiltonian
systems \citep[e.g.][]{hmm95,stoffer95,hl97,hlr01,hlw02}.
Since SPH in the absence of dissipative terms can derived from a
Hamiltonian variational principle, much of this work is applicable in the SPH
context. The primary condition for
the construction of a geometric integrator
is time-reversibility (that is, particle quantities should return to their
original values upon reversing the direction of time integration). It is fairly
straightforward to construct a reversible integrator for the SPH equations in the
case of a constant smoothing length, where the density summation is used
and where the pressure is calculated directly from the density (such that the force
evaluation uses only the particle co-ordinates). The standard leapfrog
algorithm is one such example. In general, however, the construction of a reversible
scheme is complicated by several factors. The first is the use of a variable timestep
(which immediately destroys the time-symmetry in the leapfrog scheme,
although see \citet*{hlr01} for recent progress on this).  The second complicating factor is that the reversibility condition becomes more difficult when equations with
rates of change involving the particle velocity are used (such as the thermal or
total energy equation or the continuity equation for the density). In this case
the construction of a reversible integrator for SPH
necessarily involves the calculation of derivatives involving the velocity in
separate step to the force evaluation, leading to additional computational
expense. A third complicating factor is the use of individual particle
timesteps in large SPH codes, although symplectic methods have also been constructed for this case
\citep{hlw02}.

\subsection{Courant condition}

 The timestep is determined by the Courant condition
\begin{equation}
\mathrm{dt_c} = C_{cour}\mathrm{min}\left(\frac{h}{\rv_{sig}}\right)
\label{eq:dtc}
\end{equation}
where $h = \mathrm{min}(h_a, h_b)$ and $\rv_{sig}$ is the maximum
signal velocity between particle pairs. This signal velocity is similar to that
used in the artificial dissipation terms (\S\ref{sec:av}), except that we use 
\begin{equation}
\rv_{sig} = \frac12 \left( \rv_a + \rv_b + \beta\vert \bv_{ab}\cdot \mathbf{j} \vert \right)
\end{equation}
with $\beta = 1$ when $\bv_{ab}\cdot \mathbf{j} > 0$ (ie. where the dissipation
terms are not applied). The minimum in (\ref{eq:dtc}) is taken over all particle
interactions and typically we use $C_{cour} = 0.4$.

Although this condition is sufficient for all of the simulations described here, 
in general it is necessary to pose the additional constraint from the forces
\begin{equation}
\mathrm{dt_f} = C_f \mathrm{min}\left(\frac{h_a}{\vert \mathbf{a}_a \vert}\right)^{1/2},
\end{equation}
where $\mathbf{a}_a$ is the acceleration on particle $a$ and typically $C_f = 0.25$.

\section{Numerical tests}
\label{sec:hydrotests}

\subsection{Implementation}
\label{sec:practicalissues}
Unless otherwise indicated the simulations use the density summation (\ref{eq:rhosum}),
the momentum equation (\ref{eq:sphmom}) and the energy equation in the form
(\ref{eq:sphenergy}). The numerical tests presented throughout this thesis were
implemented using a code written by the author as a testbed for MHD algorithms.

\subsubsection{Neighbour finding}
Since the code has been designed for flexibility rather than performance, we take a
simplified approach to neighbour finding using linked lists. The particles are binned
into grid cells of size $2h$ where $h$ is the maximum value of smoothing length over the
particles. Particles in a given cell then search only the adjoining cells for
contributing neighbours. This approach becomes very inefficient for a large
range in smoothing lengths such that for large simulations it is essential to use a more
effective algorithm. A natural choice is to use the tree code used in the computation of
the gravitational force \citep{hk89,bea90}.

\subsubsection{Boundary conditions}
\label{sec:ghosts}
 Boundary conditions are implemented using either ghost or fixed particles. For
reflecting boundaries, ghost particles are created which mirror the SPH particles across
the boundary. These particles are exact copies of the SPH particles in all respects
except for the velocity, which is of opposite sign on the ghost particle, producing a
repulsive force at the boundary. For periodic boundary conditions the ghosts are exact
copies of the particles at the opposite boundary. In the MHD shock tube tests considered in
\S\ref{sec:1Dtests} involving non-zero velocities at the boundaries, 
boundary conditions are implemented in one dimension by simply fixing the
properties of the 6 particles closest to each boundary. Where the initial
velocities of these particles are non-zero their positions are evolved
accordingly and a particle is removed from
the domain once it has crossed the boundary. Where the distance between the
closest particle and the boundary is more than the initial particle spacing
a new particle is introduced to the domain. Hence for inflow or outflow boundary
conditions the resolution changes throughout the simulation.

\subsection{Propagation and steepening of sound waves}
\label{sec:swavetests}
 We initially consider the propagation of linear sound waves in SPH. This test
is particularly important in the MHD case (\S\ref{sec:mwav}) since it
highlights the instability in the momentum-conserving formalism of SPMHD. In this case we
investigate the dependence of sound speed on smoothing length and the damping due to
artificial viscosity.

\subsubsection{Particle setup}
The particles are initially setup at equal separations in the domain $x =
[0,1]$ using ghost particles
(\S\ref{sec:ghosts}) to create periodic boundary conditions. The linear solution for a
travelling sound wave in the x-direction is given by
\begin{eqnarray}
\rho (x,t) & = & \rho_0 (1 + A \mathrm{ sin}(kx_a-\omega t), \\
\rv_x (x,t) & = & C_s A \mathrm{ sin}(kx_a - \omega t),
\label{eq:swave}
\end{eqnarray}
where $\omega = 2\pi C_s/\lambda$ is the angular frequency, $C_s$ is the sound speed in the undisturbed medium and $k =
2\pi/\lambda$ is the wavenumber. The initial conditions therefore correspond to $t=0$ in
the above. The perturbation in density is applied by perturbing the particles from an
initially uniform setup. We consider the one dimensional perturbation
\begin{equation}
\rho = \rho_0 [1 + \mathrm{A sin(k}x) ],
\end{equation}
where $A = D/\rho_0$ is the perturbation amplitude. The cumulative total mass in the x
direction is given by
\begin{eqnarray}
M(x) & = & \rho_0\int [1 + \mathrm{A sin (k}x) ] \mathrm{dx} \nonumber \\
& = & \rho_0 [x - \mathrm{ A cos (k}x) ]^{x}_{0},
\end{eqnarray}
such that the cumulative mass at any given point as a fraction of the total mass
is given by
\begin{equation}
\frac{M(x)}{M(x_{max})}.
\end{equation}
For equal mass particles distributed in $x=[0,x_{max}]$ the cumulative mass fraction at particle
$a$ is given by $x_a/x_{max}$ such that the particle position may be calculated
using
\begin{equation}
\frac{x_a}{x_{max}} = \frac{M(x_a)}{M(x_{max})}.
\end{equation}
Substituting the expression for $M(x)$ we have the following equation for the
particle position
\begin{equation}
\frac{x_a}{x_{max}} - \frac{x_a - \mathrm{A cos(k}x_a)}{[x_{max} - \mathrm{A cos(k}x_{max})]} = 0,
\end{equation}
which we solve iteratively using a simple Newton-Raphson rootfinder. With the uniform
particle distribution as the initial conditions this converges in one or two
iterations.

\begin{figure}
\begin{center}
\begin{turn}{270}\epsfig{file=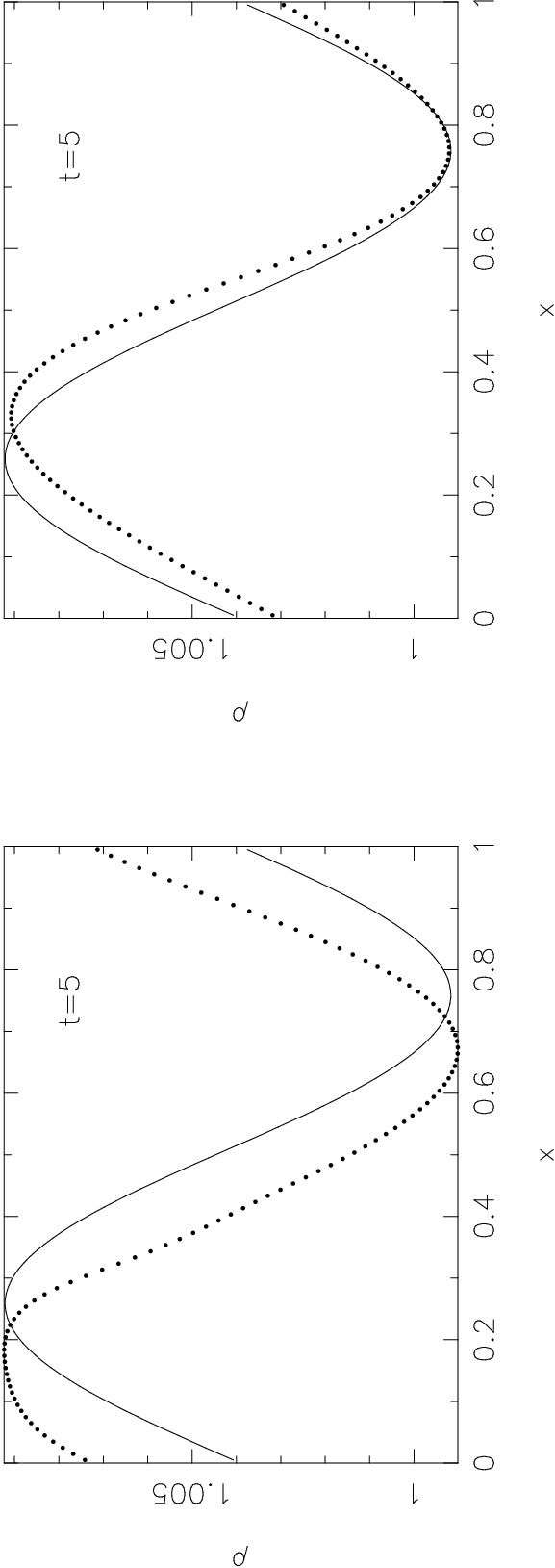,height=0.8\textwidth}\end{turn}
\caption{Representative results from the isothermal sound wave tests in one dimension using the standard cubic spline kernel
with a fixed smoothing length. The
figure on the left shows the results after 5 periods (corresponding to 5 crossings of the
computational domain) using $h = 1.5\bar{\Delta p}$.
The figure on the right shows the results using a fixed smoothing length but
with the correction from the variable smoothing length terms.}
\label{fig:isoswave_fixedh}

\begin{turn}{270}\epsfig{file=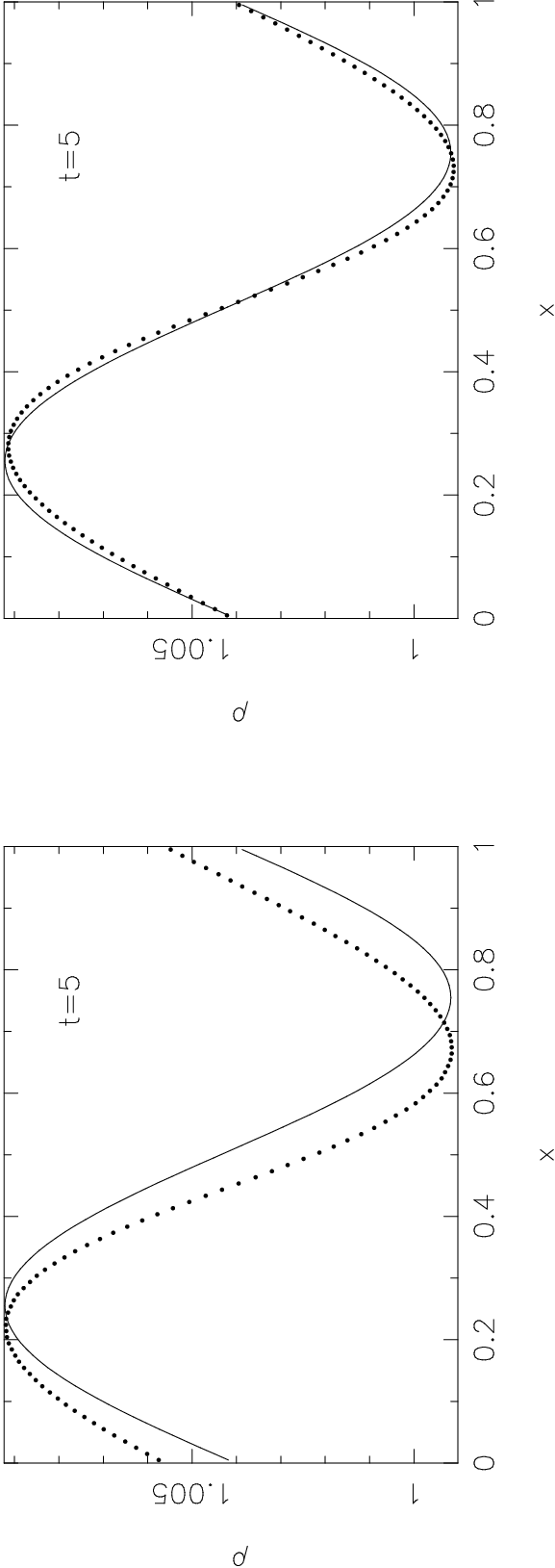,height=0.8\textwidth}\end{turn}
\caption{Representative results from the isothermal sound wave tests in one dimension using the standard cubic spline kernel
with a variable smoothing length that varies with density. The
figure on the left shows the results after 5 periods using a simple average of the kernel
gradients, whilst the figure on the right shows the results using the consistent formulation
of the variable smoothing length terms.}
\label{fig:isoswave_varh}
\end{center}
\end{figure}

\subsubsection{One dimensional tests}
 Initially we consider one dimensional, isothermal simulations using a fixed smoothing
length (for which the results of the stability analysis given in
\S\ref{sec:kernelstability} hold). The cubic spline kernel is used with $h = 1.5\Delta p$
where $\Delta p$ is the initial particle spacing. This value of smoothing length was
chosen because in Figure \ref{fig:kernelstability} the cubic spline is seen to
significantly underestimate the sound speed at this value of $h$. The simulation is setup
using 100 particles (corresponding to $k_x = 0.0628$ in Figure
\ref{fig:kernelstability}) and a wave amplitude of $0.005$ to ensure that the wave
remains essentially linear throughout the simulation.  No artificial viscosity
is used. For isothermal simulations, the pressure is calculated
directly from the density using $P = c_s^2 \rho$. The sound speed given by the SPH
simulations is estimated from the temporal spacing of minima in the total kinetic
energy of the particles.

\begin{figure}
\begin{center}
\begin{turn}{270}\epsfig{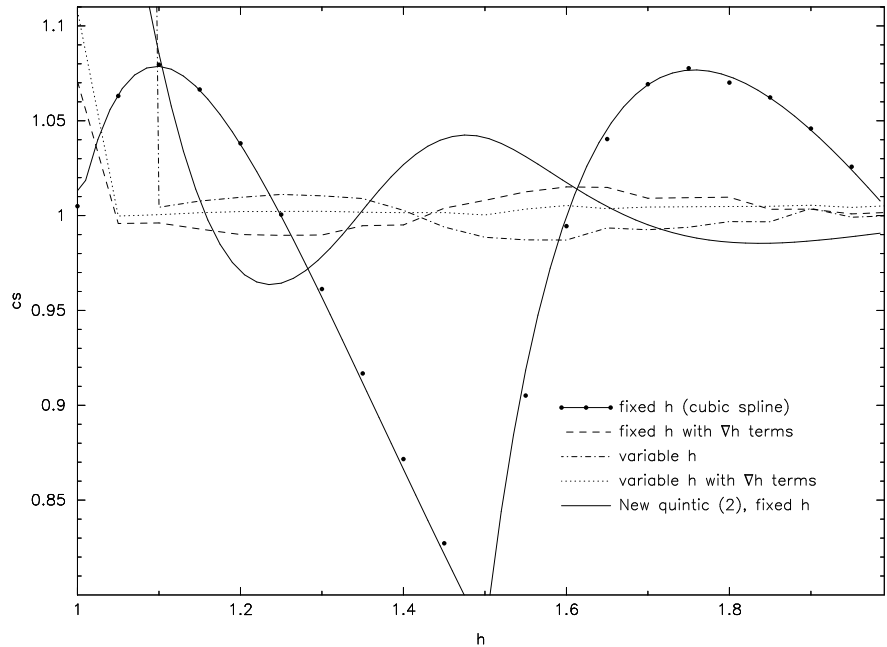}\end{turn}
\caption{Summary of the isothermal sound wave tests using 100 particles. The
numerical sound speed from the SPH simulations is shown plotted against the (mean) smoothing length in units of the
average particle spacing. Results using the cubic spline kernel with a fixed
smoothing length (solid points) may be compared with the analytic
result (solid line, under points) from the dispersion relation (\ref{eq:sphdispersion}) (this line
corresponds to $kx = 0$ in Figure \ref{fig:kernelstability}). The dashed line
gives the numerical results using the cubic spline with a fixed smoothing
length but incorporating the correction from the $\nabla h$ terms, which show much lower
phase errors. The dotted and
dot-dashed lines give numerical results using the cubic spline with a variable smoothing length with and without the
$\nabla h$ terms respectively. In both cases the results show a substantial
improvement over the fixed smoothing length case, much more so than
from the use of alternative kernels (e.g. the New Quintic (2) from
\S\ref{sec:DIYkernels}, given by the solid line).}
\label{fig:iso_swave}
\end{center}
\end{figure}

 A representative example of these simulations is given in the left hand side of Figure \ref{fig:isoswave_fixedh} after five crossings of the computational
domain. The amplitude is well maintained by the SPH scheme, however the
wave lags behind the exact solution, giving a significant phase error as expected
from the stability analysis (Figure \ref{fig:kernelstability}). The sound speed obtained from the numerical tests is plotted in
Figure \ref{fig:iso_swave} for a range of smoothing length values (solid points). In this
case the results show excellent agreement with the analytic results using the dispersion relation
(\ref{eq:sphdispersion}) given by the solid line (this line corresponds to $k_x
\approx 0$ in Figure \ref{fig:kernelstability}). We observe that, depending on the value of $h$ the numerical
sound wave can both lag and lead the exact solution (in Figure \ref{fig:iso_swave} this
corresponds to sound speeds less than or greater than unity).

 In \S\ref{sec:gradh} it was
noted that the variable smoothing length terms normalise the kernel even in the case of a
fixed smoothing length. The results of the fixed smoothing length simulation with this
correction term are shown by the dashed line in Figure \ref{fig:iso_swave}, with a
representative example given in the right hand side of Figure \ref{fig:isoswave_fixedh}. The
numerical wave speed appears much closer to the theoretical value of unity.
 
 Results using a smoothing length which varies with density according to
(\ref{eq:hevol}) are given by the dot-dashed line in Figure \ref{fig:iso_swave}, with a representative example
shown in Figure \ref{fig:isoswave_varh}. The phase error is slightly
lower than either of the fixed smoothing length cases. Including the normalisation of the kernel gradient from the variable
smoothing lengths (\S\ref{sec:gradh}) gives numerical sound speeds very close to unity
(dotted line in Figure \ref{fig:iso_swave}). A representative example of these simulations
is given in the right hand panel of Figure \ref{fig:isoswave_varh} after 5 periods. The results in this
case show excellent agreement
with the exact (linear) solution, with a small amount of steepening due to nonlinear effects.

 The results of this test indicate that, whilst alternative kernels can give slight
improvements in accuracy over the standard cubic spline (\S\ref{sec:kernelstability}), a
substantial gain in accuracy can be gained firstly by the use of a variable smoothing
length and secondly by self-consistently accounting for $\nabla h$ terms in the
formulation of the SPH equations. These terms act as a normalisation of the kernel
gradient which appear to effectively remove the dependence of the numerical sound speed on the smoothing
length value.

\subsubsection{Effects of artificial viscosity}
\begin{figure}[h]
\begin{center}
\begin{turn}{270}\epsfig{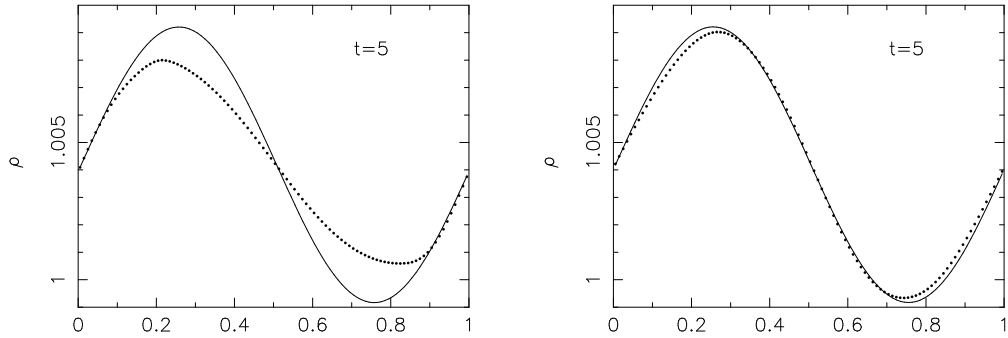}\end{turn}
\caption{(left) Isothermal sound wave with amplitude = 0.005 in one dimension with artificial viscosity
applied uniformly to particles in compression (ie. $\alpha =1$, $\beta = 2$) and (right)
applied using the viscosity switch with $\alpha_{min} = 0.1$.}
\label{fig:isoswave}
\end{center}
\end{figure}

 In the absence of any switches, the artificial viscosity is specified according
to (\ref{eq:Piab}) with $\alpha = 1$, $\beta=2$ everywhere. The results of the sound
wave propagation with artificial viscosity turned on are shown in the left panel of Figure
\ref{fig:isoswave}. After 5 crossings of the computational domain the wave is severely
damped by the artificial viscosity term. The effect is to reduce the order of the
numerical scheme since convergence to the exact solution is much slower. The results
using the artificial viscosity switch
discussed in \S\ref{sec:avlim} are shown in the right panel of Figure \ref{fig:isoswave}.
The results show good agreement with the linear solution, demonstrating that use of the artificial viscosity switch very effectively restores the numerical schemes ability to propagate small perturbations
without excessive damping.

\begin{figure}[h]
\begin{center}
\begin{turn}{270}\epsfig{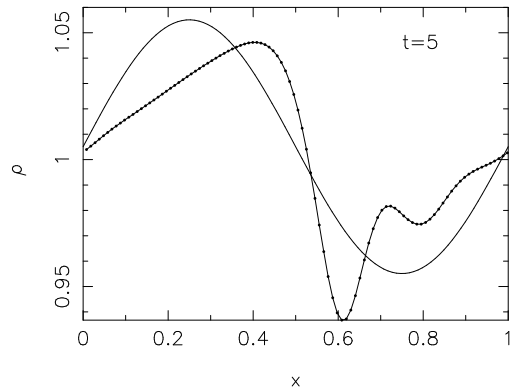}\end{turn}
\caption{Nonlinear isothermal sound wave in one dimension showing steepening to
shock. The wave profile is shown after 5 crossings of the computational domain,
corresponding to 5 periods. The initial conditions are a linear wave with amplitude 0.05 (solid
line). With
artificial viscosity applied using the switch the steepening is resolved,
although some oscillations are observed to occur ahead of the steepened wave.}
\label{fig:steepen}
\end{center}
\end{figure}

 Finally, we demonstrate the usefulness of the artificial viscosity switch by
considering the steepening of a nonlinear sound wave. In this case the initial
amplitude is 0.05 and artificial viscosity is applied using the switch. The wave
profile at $t=5$ is shown in Figure \ref{fig:steepen} and is significantly
steepened compared to the initial conditions (solid line). The use of the switch
enables the steepening to be resolved, however some oscillations are found to
occur ahead of the steepened wave.

\subsection{Sod shock tube}
\label{sec:sodshock}

 The standard shock tube test for any compressible fluid dynamics code is that
of \citet{sod78}. The problem consists of dividing the domain into two halves,
one consisting of high pressure, high density gas whilst the other is low
pressure and low density. These two portions of gas are allowed to interact
at $t=0$, resulting in a shock and rarefaction wave which propagate through the
gas. This test illustrates the shock capturing ability of the 1D code and thus
provides a good test of the artificial viscosity formalism (\S\ref{sec:av}). It
is also the basis for the MHD shock tube considered in \S\ref{sec:briowu}.  We set
up the problem using 450 SPH particles in the domain $x =[-0.5, 0.5]$. The
particles are setup with uniform masses such that the density jump is modelled by
a jump in particle separation. Initial conditions in the fluid to the left of the
origin are given by $(\rho,P,\rv_x) = [1,1,0]$ whilst 
conditions to the right are given by $(\rho,P,\rv_x)=[0.125,0.1,0]$ with
$\gamma=1.4$. The particle separation to the left of the discontinuity is $0.01$.

\begin{figure}[ht!]
\begin{center}
\begin{turn}{270}\epsfig{file=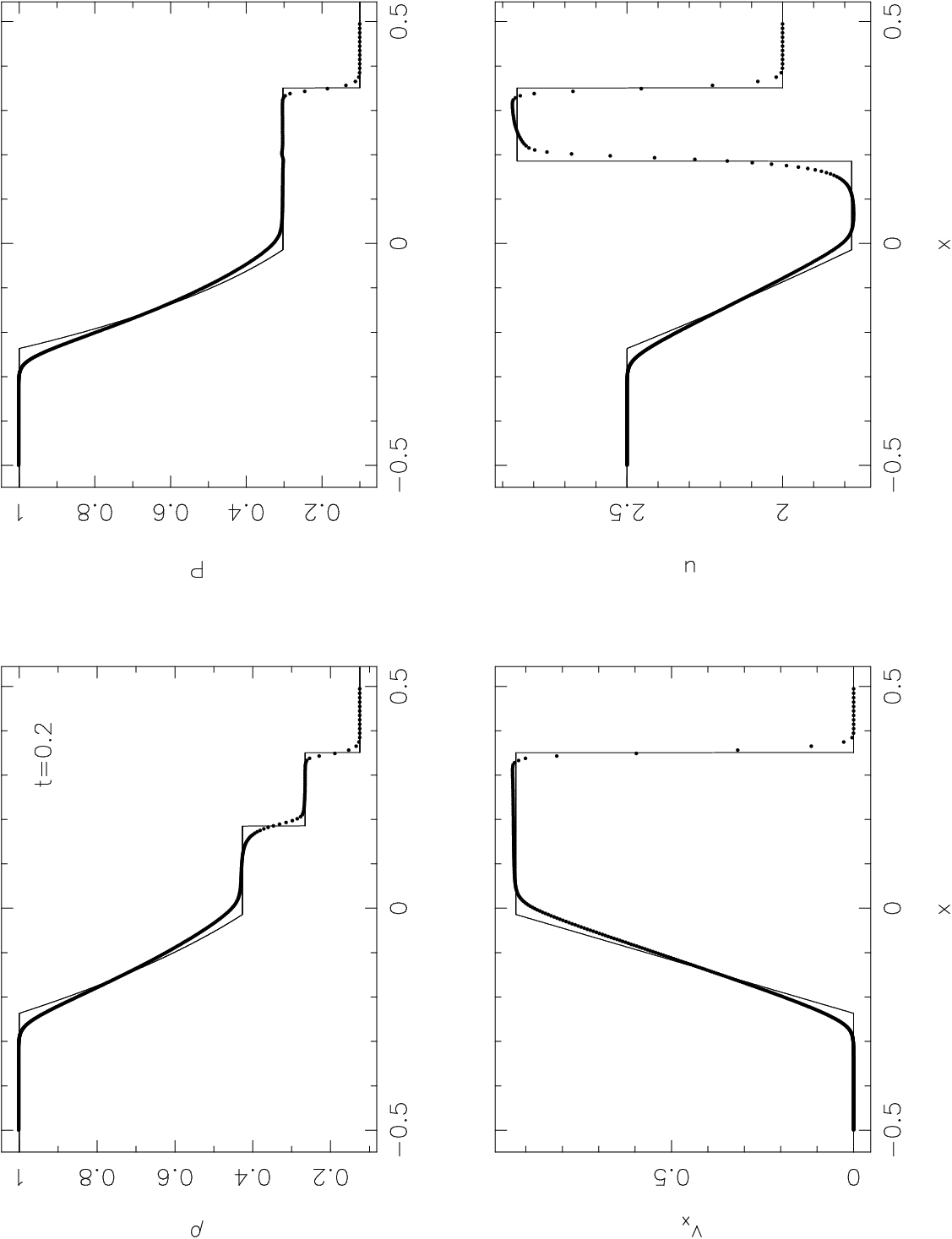,height=0.6\textwidth}\end{turn}
\caption{Results of the Sod shock tube problem in one dimension. The simulation
uses 450 particles with conditions in the fluid initially to the left of the origin
given by $(\rho,P,\rv_x) = [1,1,0]$ whilst 
conditions to the right are given by $(\rho,P,\rv_x)=[0.125,0.1,0]$ with
$\gamma=1.4$. Initial profiles of density and pressure have been smoothed and artificial viscosity is applied
uniformly. Agreement with the exact
solution (solid line) is generally good, but note the deviation from
the exact solution in the rarefaction wave due to the initial smoothing.}
\label{fig:sod_smoothed_av}

\begin{turn}{270}\epsfig{file=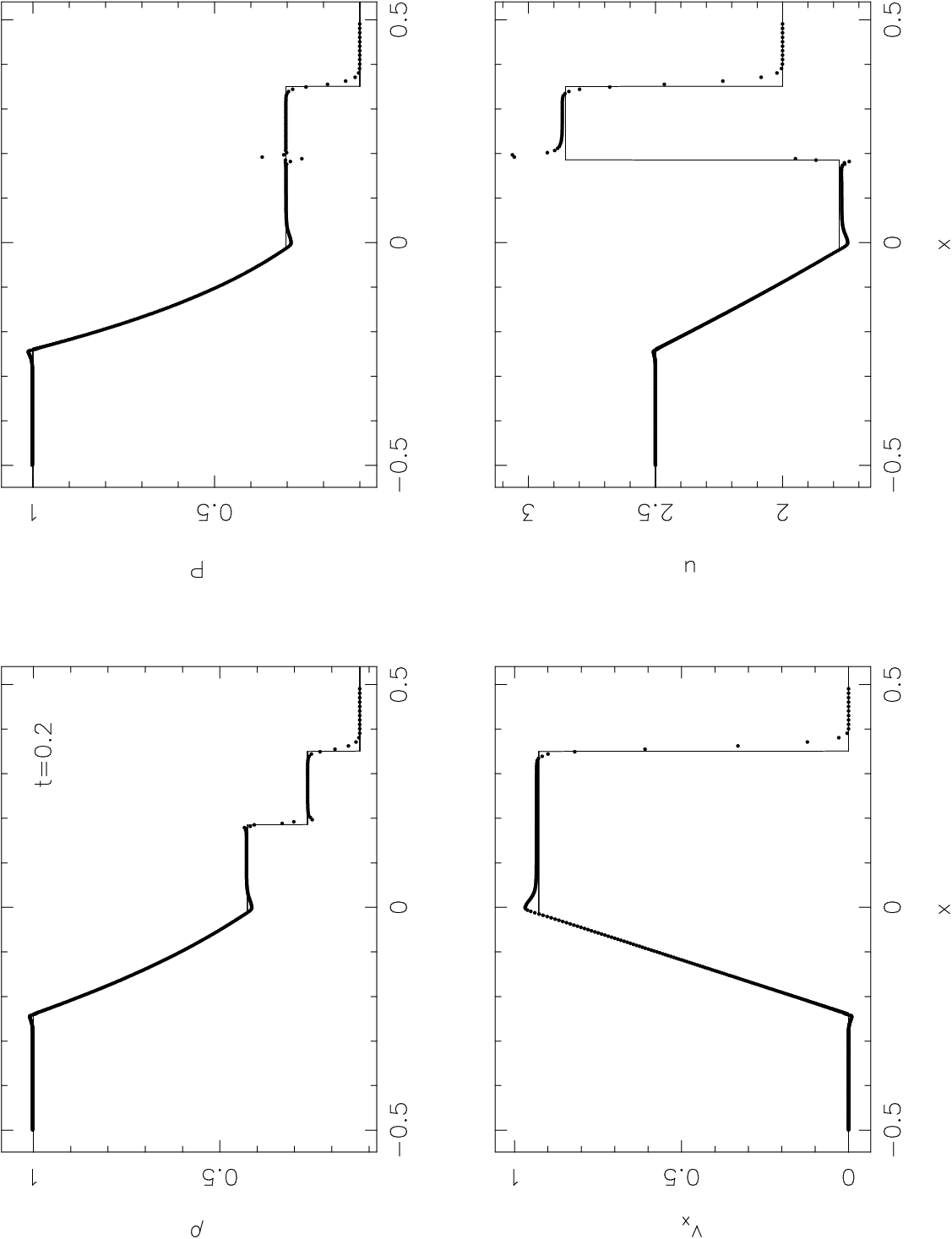,height=0.6\textwidth}\end{turn}
\caption{Results of the Sod shock tube problem using unsmoothed
(purely discontinuous) initial conditions. Artificial
viscosity has been applied uniformly whilst no artificial thermal conductivity has
been used. In the absence of any smoothing of the initial conditions the
rarefaction profile agrees well with the exact solution (solid line). The thermal
energy is observed to overshoot at the contact discontinuity. There is also a small
overshoot in velocity at the right end of the rarefaction wave.}
\label{fig:sod_unsmoothed_av}
\end{center}
\end{figure}

 Figure \ref{fig:sod_smoothed_av} shows the results of this problem at $t=0.2$. The
exact solution, calculated using the exact Riemann solver given in \citet{toro92}
is given by the solid line. In this case artificial viscosity has been applied
uniformly to particles in compression (ie. using $\alpha = 1$), whilst no
artificial thermal conductivity has been used (ie. $\alpha_u = 0$). The results are generally good although there is
significant deviation in the slope of the rarefaction wave. This can be traced
largely to the smoothing applied to the initial conditions. Following
\citet{monaghan97} (although a similar procedure is applied in many
published versions of this test), the initial discontinuities in density
and pressure were smoothed over several particles according to the rule
\begin{equation}
A = \frac{A_L + A_R e^{x/d}}{1 + e^{x/d}}
\end{equation}
where $A_L$ and $A_R$ are the uniform left and right states with respect to the
origin and $d$ is taken as half of the largest initial particle separation at
the interface (ie. the particle separation on the low density side). Where the
initial density is smoothed the particles are spaced according to the rule
\begin{equation}
\rho_a(x_{a+1} - x_{a-1}) = 2 \rho_R \Delta_R
\end{equation}
where $\Delta_R$ is the particle spacing to the far right of the origin with
density $\rho_R$. Note that initial smoothing lengths are set according to the
rule $h \propto 1/\rho$ and are therefore also smoothed. Where the total energy
$\hat{\epsilon}$ is integrated we smooth the basic variable $u$ construct the total
energy from the sum of the kinetic and internal energies. 

\begin{figure}[ht!]
\begin{center}
\begin{turn}{270}\epsfig{file=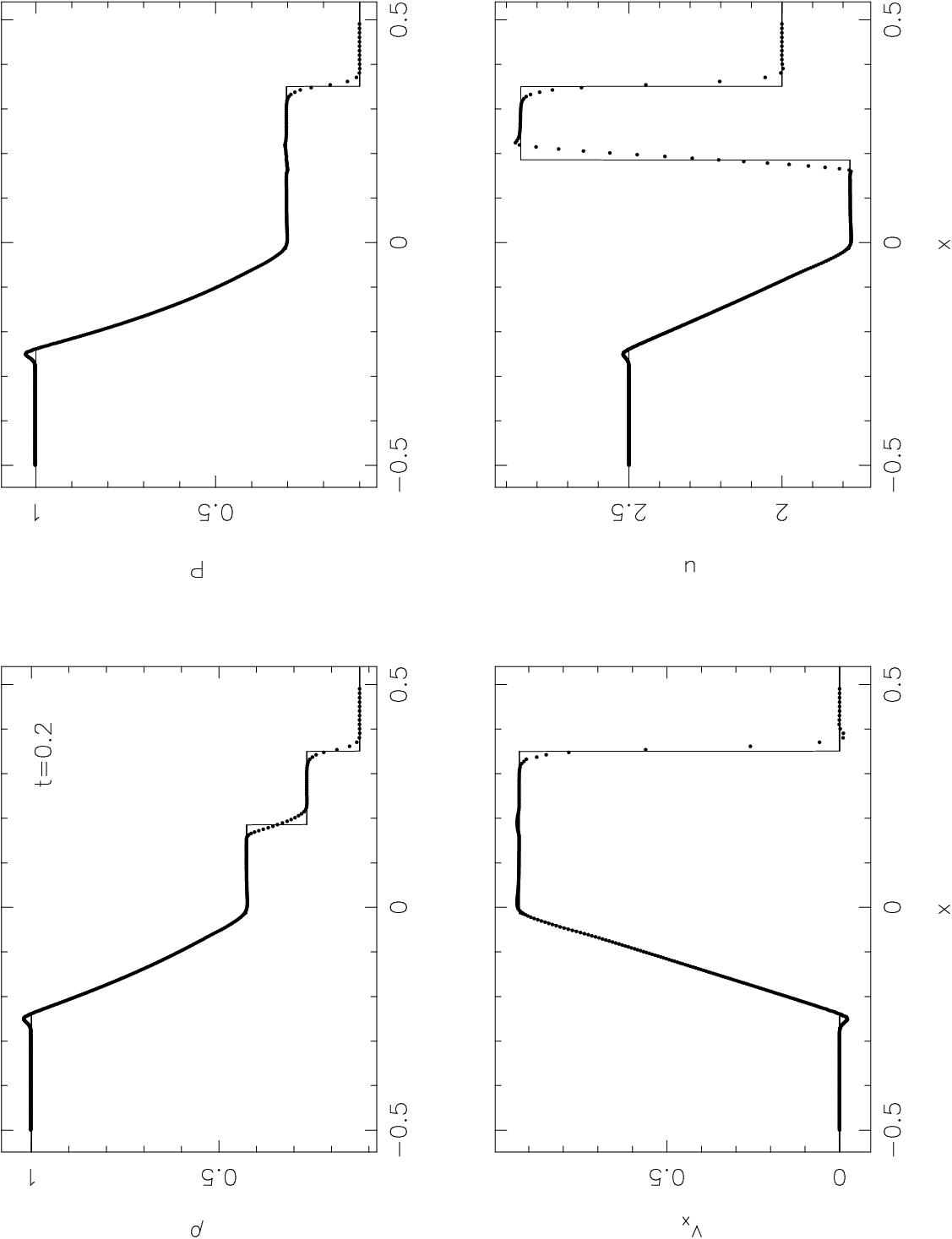,height=0.6\textwidth}\end{turn}
\caption{Results of the Sod shock tube problem using unsmoothed
initial conditions and applying a small amount of artificial
thermal conductivity using the switch described in \S\ref{sec:condswitch}. Artificial viscosity is applied uniformly. The overshoot in the thermal energy observed
in Figure \ref{fig:sod_unsmoothed_av} is corrected for by the smoothing of the
contact discontinuity produced by the thermal conductivity term. The variable
smoothing length terms have also been used in this case, although results are
similar with a simple average of the particle kernels.}
\label{fig:sod_unsmoothed_gradh_cond}

\begin{turn}{270}\epsfig{file=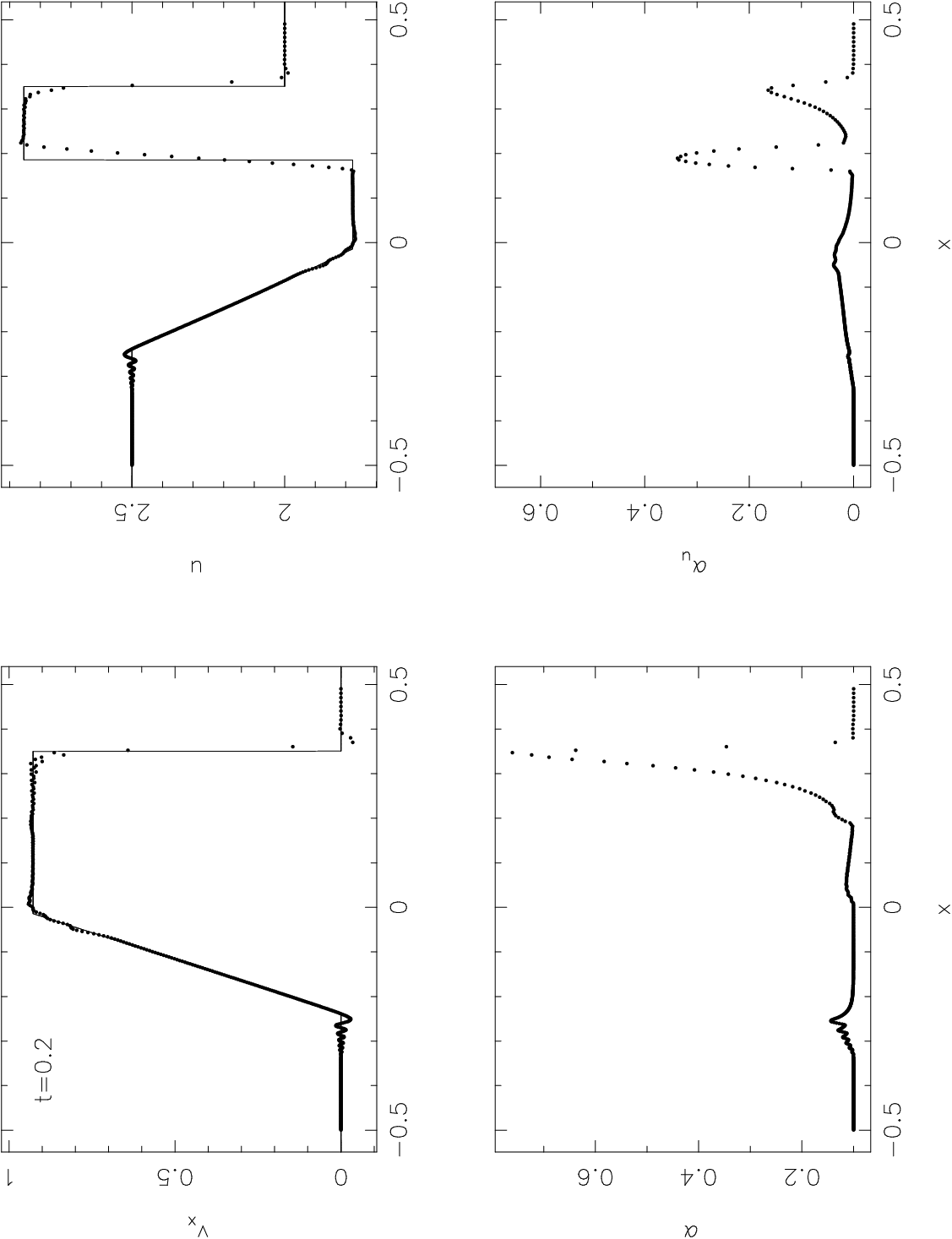,height=0.6\textwidth}\end{turn}
\caption{Velocity and thermal energy profiles (top row) in the Sod shock tube problem using unsmoothed
initial conditions and where both artificial viscosity and
thermal conductivity are applied using the switches discussed in \S\ref{sec:avlim}.
The bottom row shows the time-varying co-efficients $\alpha$ and $\alpha_u$ of the viscosity and thermal
conductivity respectively. With the unsmoothed initial conditions and the viscosity
switch there is a slight oscillation in the velocity profile at the head of the
rarefaction wave. The variable smoothing length terms have also been used in this
case.}
\label{fig:sod_switches}
\end{center}
\end{figure}

 Such smoothing of the initial conditions can be avoided altogether if the density summation (\ref{eq:rhosum}) is used,
particularly if the smoothing length is updated self-consistently with the density.
The results of this problem using unsmoothed initial conditions are shown in Figure
\ref{fig:sod_unsmoothed_av}. The artificial viscosity is applied uniformly whilst
no artificial thermal conductivity has been used. In this case the rarefaction
profile agrees extremely well with the exact solution (solid line). The unsmoothed
initial conditions highlight the need for artificial thermal conductivity since the
thermal energy overshoots at the contact discontinuity with a resulting glitch in the pressure profile. The gradient in
thermal energy at the shock front does not show this effect due to the smoothing of
the shock by the artificial viscosity term. The results of this test with a small
amount of artificial thermal conductivity applied using the switch discussed in
\S\ref{sec:condswitch} are shown in Figure \ref{fig:sod_unsmoothed_gradh_cond}. The variable
smoothing length terms have also been used in this case, although results are
similar with a simple average of the kernel gradients in the force equation
(\ref{eq:sphmom}). The contact discontinuity is smoothed over several smoothing
lengths by the thermal conductivity term, removing the overshoot in the thermal
energy. The resulting profiles compare extremely well with the exact solution
(solid line).

 Finally, the results of this test where both the artificial viscosity
and conductivity are controlled using the switches described in \S\ref{sec:avlim}
are shown in Figure \ref{fig:sod_switches}. The top row shows the velocity and
thermal energy profiles compared with the exact solution (solid line), whilst the
bottom row shows the time-varying co-efficients $\alpha$ and $\alpha_u$ of the viscosity and thermal
conductivity respectively. With the unsmoothed initial conditions and the viscosity
switch there is a slight oscillation in the velocity profile at the head
of the rarefaction wave. The variable smoothing length terms have been used in this
case involving the consistent update of the smoothing length with density (\S\ref{sec:gradh}).
If a simple average of the kernel gradients is used instead the oscillations in the rarefaction wave are still present but slightly less
pronounced. In effect, the iterations of density and
smoothing length make the scheme much more sensitive to small perturbations, since
a small change in the smoothing length will be reflected in the density profile
and vice-versa. This means that structures in the simulation are in general
better resolved and is clearly advantageous. However alsos mean that
small errors in the density evolution are amplified where they may otherwise
have been smoothed out by the numerical scheme.

\subsection{Blast wave} \label{sec:blast}
 In this test we consider a more extreme version of the shock tube test considered
previously. In this problem the initial conditions in the fluid to the left of the
origin are given by $(\rho,P,\rv_x) = [1,1000,0]$ whilst 
conditions to the right are given by $(\rho,P,\rv_x)=[1,0.1,0]$ with
$\gamma=1.4$. The $10^4$ pressure ratio across the initial discontinuity results in
a strong blast wave which propagates into the fluid to the right of the origin.
The velocity of the contact discontinuity is very close to that of the shock,
producing a sharp density spike behind the shock front. This test therefore
presents a demanding benchmark for any numerical code. 

\begin{figure}[h!]
\begin{center}
\begin{turn}{270}\epsfig{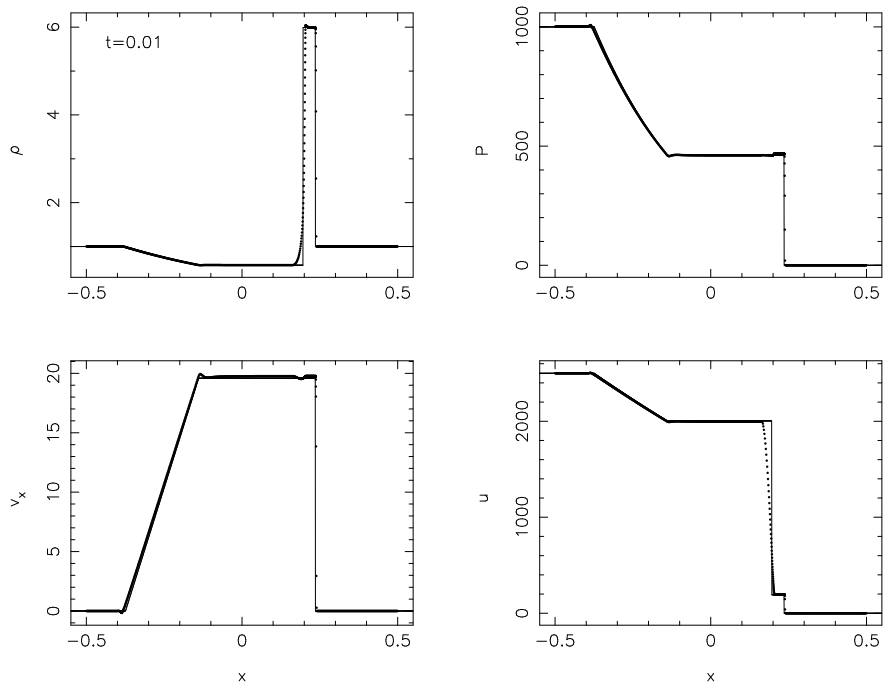}\end{turn}
\caption{Results of the one dimensional blast wave test at $t=0.01$. Conditions in the fluid initially to the left of the origin
given by $(\rho,P,\rv_x) = [1,1000,0]$ whilst 
conditions to the right are given by $(\rho,P,\rv_x)=[1,0.1,0]$ with
$\gamma=1.4$. 1000 particles have been used with no smoothing of the initial conditions.
The agreement with the exact solution (solid line) is excellent. The contact
discontinuity is spread sufficiently by the artificial thermal conductivity to be
resolved accurately. In this simulation the density summation and the average of the kernel gradients has
been used.}
\label{fig:blast}
\end{center}
\end{figure}

The results of this test at $t=0.01$ are shown in Figure \ref{fig:blast}. The
agreement with the exact solution (solid line) is excellent.  In this simulation the density summation and the average of the kernel gradients has
been used and the artificial viscosity is applied using the viscosity switch. The SPH results may be compared with those given in \citet{monaghan97}.
Although we use the same formulation of the dissipative terms as in
\citet{monaghan97}, in that paper the artificial thermal conductivity is applied only for
particles in compression, resulting in a need to smooth the initial discontinuity
in the pressure. With the thermal conductivity term applied using the switch the contact
discontinuity is spread sufficiently in order to be resolved accurately and
smoothing of the initial conditions is therefore unnecessary.  In the SPH
solution given by \citet{monaghan97} the spike in density is observed to
overshoot the exact solution, which is not observed in this case. This is
due to the use of the density summation (\ref{eq:rhosum}) rather than evolving the continuity
equation (\ref{eq:sphcty}) as in \citet{monaghan97}. Use of the continuity
equation is more efficient since it does not require an extra pass over the
particles in order to calculate the density. Using alternative formulations of
the pressure term in the momentum equation (e.g. using equation (\ref{eq:genmom}) with $\sigma =
1$) gives similar results (although the \citet{hk89} formulation
(\ref{eq:hkmom}) appears to produce negative pressures on this problem). Using
the consistent alternative formulations of the continuity equation,  however, appears to worsen
the overshoot observed in the density spike compared to the usual continuity
equation (for example in the $\sigma = 1$ case, the density spike overshoots to
$\rho_{max} \approx 10$ when the continuity equation (\ref{eq:altcty}) is used).

\subsection{Cartesian shear flows}
\label{sec:shearflows}
 In a recent paper \citet{ii02} (hereafter II02) have suggested that SPH gives particularly poor
results on problems involving significant amounts of shear. The simplest test considered
by II02 is a Cartesian shear flow. The setup is a two dimensional, uniform density $\rho
= 1$ box in the
domain $0 \le x \le 1$ and $0 \le y \le 1$ with a shear velocity field $\rv_x = 0, \rv_y =
\sin{(2\pi x)}$ and periodic boundary conditions in the $x-$ and $y-$ directions. In general such flows are known (at least in the incompressible case) to be unstable to Kelvin-Helmholtz
instabilities at the inflection point in the velocity profile \citep[e.g.][]{drazinreid}.
However, a straightforward stability analysis of this flow demonstrates that it is indeed
stable to small perturbations in the $x-$direction (note, however that the application of viscosity
can significantly affect the stability properties for these types of problems).

\begin{figure}[h]
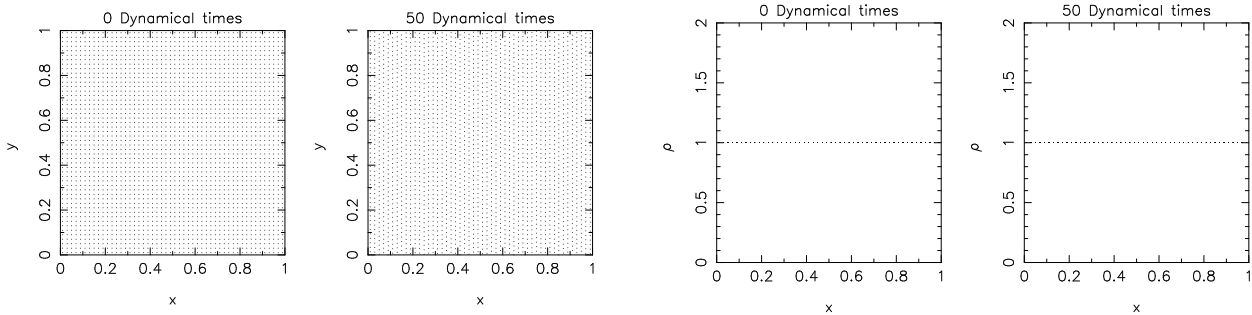

\centering
\begin{minipage}{0.47\textwidth}
\begin{turn}{270}\epsfig{file=shearcs0_part.ps,height=1.0\textwidth}\end{turn}
\end{minipage}\hspace{0.05\textwidth}
\begin{minipage}{0.47\textwidth}
\begin{turn}{270}\epsfig{file=shearcs0_dens.ps,height=1.0\textwidth}\end{turn}
\end{minipage}
\caption{Particle positions (left) and density evolution (right) in the pressure-free
Cartesian shear flow test with shear velocity field $\rv_x = 0, \rv_y = \sin{(2\pi x)}$. The amplitude of the density error is extremely small ($\Delta \rho \approx
10^{-3}\rho$)}
\label{fig:cartshearcs0}
\end{figure}

\begin{figure}[h]
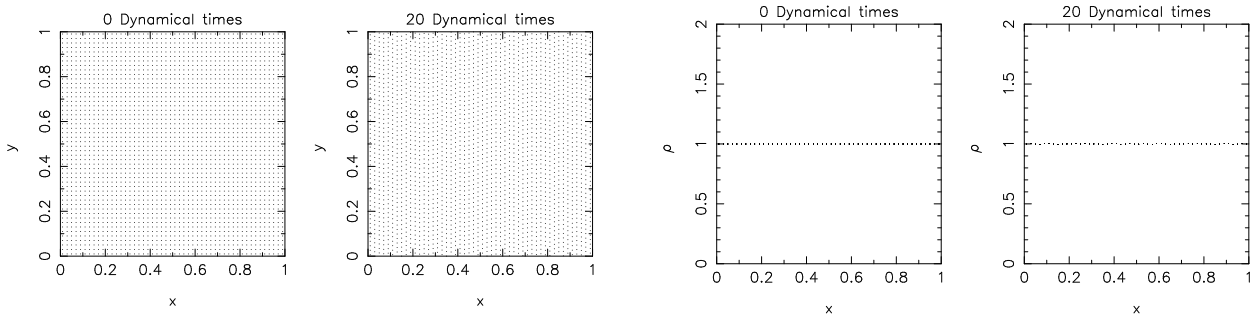

\centering
\begin{minipage}{0.47\textwidth}
\begin{turn}{270}\epsfig{file=shearcs005_part.ps,height=1.0\textwidth}\end{turn}
\end{minipage}\hspace{0.05\textwidth}
\begin{minipage}{0.47\textwidth}
\begin{turn}{270}\epsfig{file=shearcs005_dens.ps,height=1.0\textwidth}\end{turn}
\end{minipage}
\caption{Particle positions (left) and density evolution (right) in the Cartesian shearing box test with sound speed
$c_{s0} = 0.05$ and shear velocity field $\rv_x = 0, \rv_y = \sin{(2\pi x)}$. The amplitude of the density error is very small ($\Delta \rho \approx
10^{-2}\rho$)}
\label{fig:cartshearcs005}
\end{figure}

 We setup the problem using 2500 (50 x 50) particles initially arranged on a cubic
lattice. The smoothing length we use is set according to
\begin{equation}
h = \eta \left(\frac{m}{\rho}\right)^\frac12,
\label{eq:heta}
\end{equation}
where we use $\eta = 1.2$, resulting in an initially uniform value of $h = 0.024$. The smoothing length is allowed to
change with density according to (\ref{eq:hevol}), although this has little effect since
the density remains close uniform throughout the simulation. The equation of state is isothermal such that the pressure is given in terms of
the density via $P = c_s^2 \rho$. As in II02, we consider both the pressure-free case ($c_s = 0$)
and also using $c_s = 0.05$, in both cases using no artificial viscosity. The results for the pressure-free case are shown in Figure
\ref{fig:cartshearcs0}. After 50 dynamical times (defined as one crossing of the computational domain at the highest
velocity, ie. in this case $t_{dyn} = 1$) the density remains extremely close to
uniform ($\Delta \rho \approx 10^{-3}\rho$) and the particle positions remain ordered. Results in II02 show large errors ($\Delta \rho /\rho
\gtrsim \rho$) in the density in less than 1 dynamical time. Similar results are obtained
in the $c_s = 0.05$ case, shown after 20 dynamical times in Figure
\ref{fig:cartshearcs005}. Again, the amplitude of the density error is very small ($\Delta \rho \approx
10^{-2}\rho$). Some disruption in the particle
distribution is observed to occur at later times, however in the absence of any artificial viscosity
small compressible modes are not damped in any way and in the absence of a high accuracy
timestepping algorithm such disorder might reasonably be expected. Also it is well known
that the particles initially arranged on a cubic lattice will eventually move off the
lattice and settle to a more isotropic close packed distribution (e.g. \citealt{morrisphd}).

 The question is, therefore: Why do the results obtained in II02 show so much error in the
density evolution? The
major factor appears to be the particle setup. The details of the particle setup are not
given in II02, however by inspection of their figures it appears that the particles are
arranged in a quasi-random fashion. The density errors observed in their paper may
therefore be an amplification (by the shear flow) of initial perturbations in the
density distribution due to the particle setup. A second contributing factor is that the
value of smoothing length used by II02 is very low (they use $\eta = 1$ in
equation (\ref{eq:heta}), whereas typical values for $\eta$ lie in the
range $1.1-1.2$ in most multidimensional SPH implementations). However, even with their
choice of smoothing length $h=1.0 (m/\rho)^\frac12$, we still find that the density
remains essentially constant. 

\subsection{Toy stars}
\label{sec:toystar}

 A disadvantage of many of the test problems found in the astrophysical fluid dynamics
literature is that, being designed for grid-based codes, they all involve some
kind of boundary condition. For codes designed ultimately to simulate
self-gravitating gas it is useful to have benchmarks based on a finite system.
Secondly simple, exact, nonlinear solutions to the equations of hydrodynamics are
few and far between, and this even more so in the case of magnetohydrodynamics.
For this reason we investigate benchmarks based on a simple class of exact
solutions which we call `Toy Stars'. The equations of hydrodynamics are modified
by the addition of a linear force term which is proportional to the co-ordinates
(which means that the particles move in a paraboloidal potential centred on the
origin). The one dimensional equation of motion is given by
\begin{equation}
\frac{d\rv}{dt} = - \frac{1}{\rho} \frac{\partial P}{\partial x}  - \Omega^2 x,
\label{eq:toystarmom}
\end{equation}
where $\Omega$ is the angular frequency. In the following we rescale the equations in
units such that $\Omega^2 = 1$. The toy star force has many interesting properties
and was even considered by Newton as an example of the simplest many-body force. The toy star equations with
$\gamma = 2$ are also identical in form to the shallow water equations.

 Assuming a polytropic equation of state (ie. $P = K \rho^\gamma$) with constant
of proportionality $K = 1/4$ and $\gamma = 2$, the Toy Star static structure at
equilibrium is easily derived from (\ref{eq:toystarmom}) as
\begin{equation}
\rho = \rho_0 (1 - x^2) \label{eq:staticrho}
\end{equation}

\begin{figure}
\begin{center}
\begin{turn}{270}\epsfig{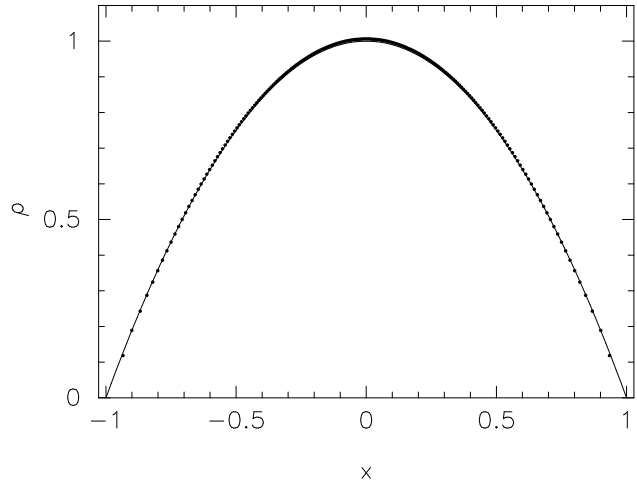}\end{turn}
\caption{Toy star static structure. 200 SPH particles are set up in an initially uniform distribution
along the x axis and allowed to evolve under the influence of the linear force.
The SPH particles are shown by the solid points after damping to an equilibrium
distribution. The agreement with the exact quadratic ($\rho = 1-x^2$) solution (solid line) is extremely
good.}
\label{fig:static}
\end{center}
\end{figure}

 In this thesis we will
simply consider the most interesting toy star problem which is the calculation of
the fundamental oscillatory mode since it
turns out to be an exact, non-linear solution.  However, a perturbation analysis can be used to derive linear solutions to the Toy Star
equations which also present interesting benchmarks for numerical codes. An investigation
of the linear modes using SPH, together with a detailed comparison of the oscillation
frequencies with the linear solution is given in \citet{mp04}.
The non-linear solution for arbitrary $\gamma$ may be derived by considering velocity perturbations in the form
\begin{equation}
\rv = A(t) x,
\end{equation}
where the density is given by
\begin{equation}
\rho^{\gamma -1} = H(t) - C(t) x^2.
\end{equation}
The exact solution \citep{mp04} for the parameters A, H and C is given in terms of the ordinary
differential equations
\begin{eqnarray}
\dot{H} & = & -AH(\gamma -1), \label{eq:toystarexact1} \\
\dot{A} & = & \frac{2K \gamma}{\gamma -1} C - 1 - A^2 \\
\dot{C} & = & -AC(1+ \gamma). \label{eq:toystarexact3}
\end{eqnarray}
which can be solved numerically with ease. The relation
\begin{equation}
A^2 = -1 - \frac{2 \sigma C}{\gamma -1} + kC^{\frac{2}{\gamma +1}},
\label{eq:kconst}
\end{equation}
where $k$ is a constant which is determined from the initial values of $A$ and $C$. The
exact solution equations (\ref{eq:toystarexact1})-(\ref{eq:toystarexact3}) take particularly
simple forms for the case $\gamma = 2$.

\subsubsection{Static structure}
\label{sec:static}
 The simplest test with the toy star is to verify the static structure. We
setup 200 SPH particles equally spaced along the x axis with $x = [-1,1]$ with
zero initial velocity and a total mass $M = 4/3$. The particles are then allowed
to evolve under the influence of the linear force, with the velocities damped
using the artificial viscosity. The particle distribution at equilibrium is shown in Figure \ref{fig:static} and
shows extremely good agreement with the exact solution (eq. \ref{eq:staticrho}).

\subsubsection{Non linear test cases}
\label{sec:nonlineartests}
\begin{figure*}
\begin{center}
\begin{turn}{270}\epsfig{file=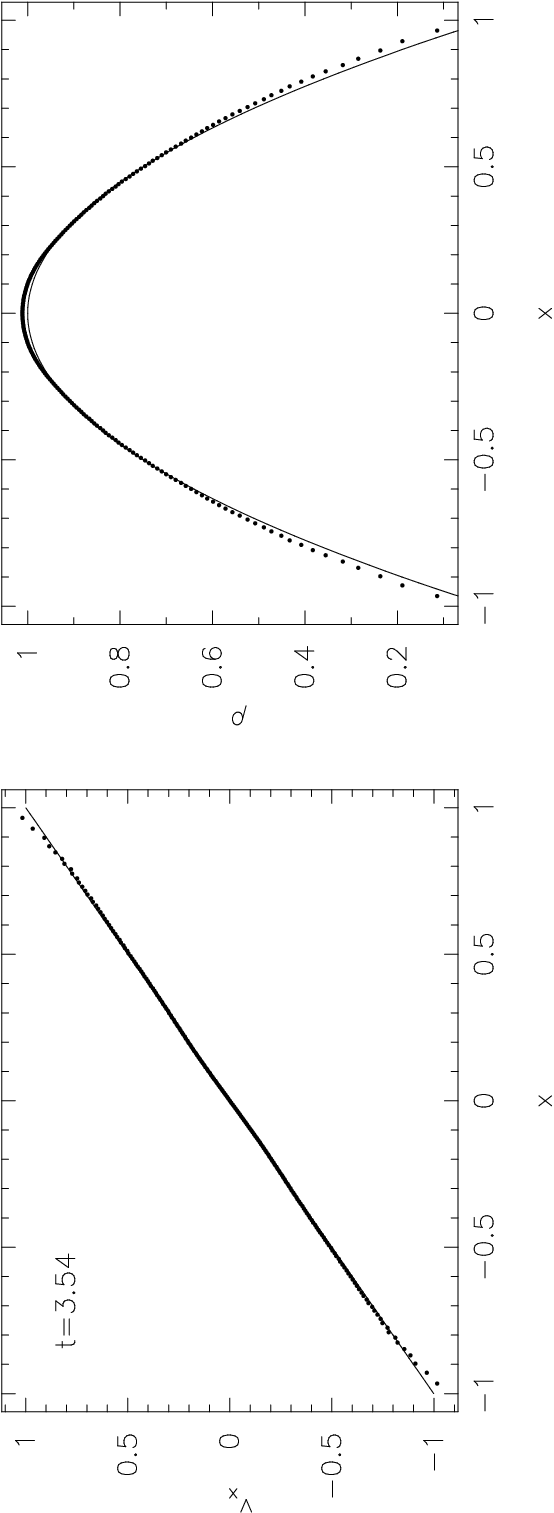,height=\textwidth}\end{turn}
\caption{Results of the SPH non linear Toy star simulation with $\gamma = 2$ and
initial conditions $\rv=x$, $\rho = 1-x^2$ (ie. $A=C=H=1$). Velocity and density profiles are shown after approximately one
oscillation period, with the SPH particles indicated by the solid points and the
exact solution by the solid line in each case. Equal mass particles are used with a variable initial
separation.}
\label{fig:equalmass1}

\begin{turn}{270}\epsfig{file=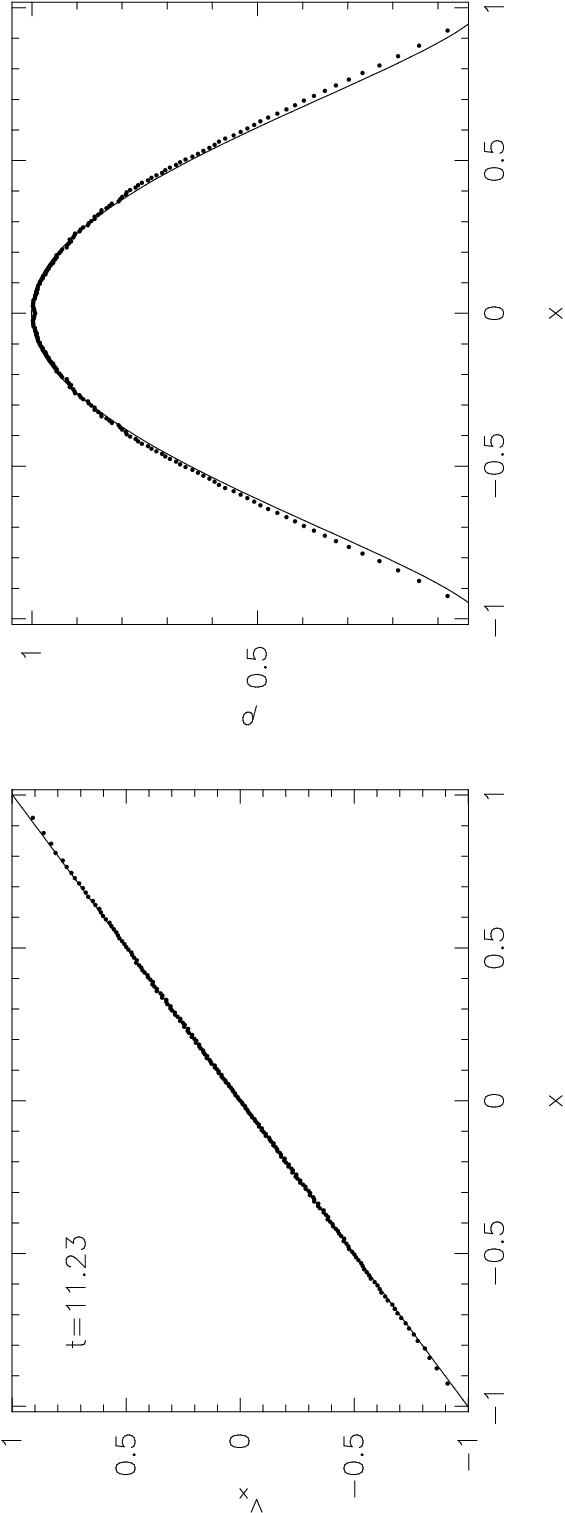,height=\textwidth}\end{turn}
\caption{Results of the SPH non linear Toy star simulation with $\gamma = 5/3$
and initial conditions $\rv=x$, $\rho = (1-x^2)^{3/2}$ (ie. $A=C=H=1$ with $\gamma = 5/3$). Velocity and density profiles are shown after approximately three
oscillation periods and the exact solution is given by the solid line.}
\label{fig:gam53}
\end{center}
\end{figure*}

For the non-linear tests the one dimensional Toy star is initially set up using 200 equal mass particles distributed along the x
axis. Although in principle we could use the particle distribution obtained in the
previous test as the initial conditions, it is simpler just to space the
particles according to the static density profile (\ref{eq:staticrho}). The SPH equations are implemented using the summation over particles to
calculate the density and the usual momentum equation with the linear force
subtracted. The equation of state is specified by using $P = K \rho^\gamma$,
where for the cases shown we set $K=1/4$. The smoothing length is allowed to vary with the particle density, where we take simple averages of kernel quantities in
the SPH equations in order to conserve momentum.

 The exact (non-linear) solution is obtained by numerical integration of equations
(\ref{eq:toystarexact1})-(\ref{eq:toystarexact3}) using a simple improved Euler method. We use the
condition (\ref{eq:kconst}) as a check
on the quality of this integration by evaluating the constant $k$, which should
remain close to its initial value.

 Results for the case where initially $A=C=H=1$ (and therefore $k =
 4$) are shown in figure \ref{fig:equalmass1} at $t=3.54$ (corresponding to approximately one oscillation
period) alongside the exact solution shown by the solid lines. No artificial
viscosity is applied in this case. The agreement with the exact solution is
excellent. Note that the sound speed in this case is $C_s = 1/\sqrt{2}$ such
that using the parameter $A=1$ results in supersonic velocities at the edges of
the star (the solution is therefore highly non-linear).

 Figure \ref{fig:gam53} shows the SPH results for a simulation with $\gamma =
5/3$ and the same initial parameters as Figure
\ref{fig:equalmass1}. Velocity and density profiles are shown at
time $t=11.23$ corresponding to approximately three oscillation periods. No
artificial viscosity is used. The agreement with the exact solution (solid
lines) is again extremely good.

 Results of simulations with artificial viscosity turned on are similar, although with a small damping of the kinetic energy over time.

\section{Summary}
 In this chapter we have thoroughly reviewed the SPH algorithm. Alternatives to
the standard cubic spline kernel were investigated in \S\ref{sec:kernels} and
\S\ref{sec:DIYkernels}, on the basis of their stability properties. Higher order
spline kernels giving closer approximations to the Gaussian were found to give
better stability properties although at the price of an increase in
computational expense due to the greater number of contributing neighbours. The
possibility of constructing kernels with better
stability properties based on smoother splines but retaining compact support of size $2h$ was
investigated, with good results for smoothing lengths $h \gtrsim
1.1$ (in units of the average particle spacing). However, the gain in accuracy from the use of these
alternative kernels is very minor compared to the substantial improvements in accuracy
gained by the incorporation of the variable smoothing length terms (\S\ref{sec:gradh}) 

 The discrete equations of SPH were formulated self-consistently from a variational
principle in \S\ref{sec:fluideqs}, leading naturally to equations which explicitly conserve momentum, angular
momentum and energy. Artificial dissipation terms used to capture shocks were then
discussed, where in \S\ref{sec:condswitch} a new switch to control the application of
artificial thermal conductivity was considered (the importance of which is highlighted in
the numerical tests described in \S\ref{sec:hydrotests}). The consistent formulation of the
SPH equations incorporating a variable smoothing length was discussed in
\S\ref{sec:gradh}, which are shown to lead to increased accuracy in a wide range of
problems (including linear waves (\S\ref{sec:swavetests}), shock tubes
(\S\ref{sec:sodshock}), Cartesian shear flows (\S\ref{sec:shearflows}) and
toy stars (\S\ref{sec:toystar})). It was shown in \S\ref{sec:altforms} that consistent
formulations of SPH when alternative formulations of the momentum equation are used can
be derived from a variational principle by modifying the form of the continuity equation.
Various timestepping algorithms were discussed in \S\ref{sec:timestep}, particularly the
need to perform a separate pass over the particles to compute derivatives involving the
velocity for a reversible integration of the SPH equations. Finally several numerical tests were
presented. 

The linear sound wave tests (\ref{sec:swavetests}) demonstrated a phase error
in the SPH simulation of sound waves dependent on the value of the smoothing length and
related to the use of kernels with compact support. This phase error was shown to be
largely corrected for by allowing the smoothing length to vary with density and
self-consistently accounting for the extra terms which arise in the SPH equations. Also
the damping of small perturbations induced by the artificial viscosity term was found to
be significantly reduced by use of the artificial viscosity switch described in
\S\ref{sec:avlim}. In the second test problem, the standard shock tube test of \citet{sod78}, the
importance of applying a small amount of artificial thermal conductivity was
highlighted, which avoids the need to artificially smooth the initial conditions of such
problems. The SPH algorithm was also shown to give good results on a more extreme version
of this test (\S\ref{sec:blast}). Thirdly (\S\ref{sec:shearflows}), the Cartesian shear flow tests given by
\citet{ii02} were examined, demonstrating that SPH gives good results on this problem for
uniform particle setups and does not show the large errors encountered by these authors.
Finally, the SPH algorithm was tested against several exact, non-linear solutions derived for systems of particles,
known as `Toy Stars' and was shown to give results in excellent agreement with
theory.

\include{c4}

\include{c5}

\include{c6}

\backmatter

\appendix

\include{c2app}

\chapter{SPH stability analysis}
\label{sec:sphstability} 
 In this appendix we perform a stability analysis of the standard SPH formalism derived
in \S\ref{sec:fluideqs}. Since the SPH equations were derived directly from a variational principle, the
linearised equations may be derived from a second order perturbation to the
Lagrangian (\ref{eq:Lsph}), given by
\begin{equation}
\delta L = \smb \left[ \frac{1}{2}\rv_b^2 - \delta\rho_b\frac{du_b}{d\rho_b} -
\frac{(\delta\rho_b)^2}{2}\frac{d^2 u_b}{d\rho_b^2} \right]\label{eq:perturb}
\end{equation} 
where the perturbation to $\rho$ is to second order in the second term and to first
order in the third term. The density perturbation is given by a perturbation of the
SPH summation (\ref{eq:rhosum}), which to second order is given by\footnote{Note that the first order term may be decoded into continuum form to give the usual expression
\begin{equation}
\delta\rho = -\rho_{0}\nabla\cdot(\delta\mathbf{r}) \nonumber
\end{equation}
where $\rho_{0}$ refers to the unperturbed quantity.}
\begin{equation}
\delta\rho_{a} = \smb \delta x_{ab} \pder{W_{ab}}{x_a} + \smb\frac{(\delta
x_{ab})^2}{2} \pder{^2 W_{ab}}{x_a^2}
\label{eq:o2deltarho}
\end{equation}
 The derivatives of the thermal energy with respect to density follow from the
first law of thermodynamics, ie.
\begin{equation}
\frac{du}{d\rho} = \frac{P}{\rho^2}, \hspace{2cm} \frac{d^2 u}{d\rho^2}
= \frac{d}{d\rho}\left( \frac{P}{\rho^2} \right) = \frac{c_s^2}{\rho^2} - \frac{2P}{\rho^3} \nonumber
\end{equation}
The Lagrangian perturbed to second order is therefore
\begin{equation}
\delta L = \smb \left[ \frac{1}{2}\rv_b^2 - \frac{P_b}{\rho_b^2}\smc\frac{(\delta
x_{bc})^2}{2} \pder{^2 W_{bc}}{x_a^2}
- \frac{(\delta\rho_b)^2}{2\rho_b^2}\left(c_s^2 - \frac{2P_b}{\rho_b} \right) \right]
\end{equation}

The perturbed momentum equation is given by using the perturbed Euler-Lagrange equation,
\begin{equation}
\frac{d}{dt}\left(\pder{L}{{\rm v}_a}\right) - \pder{L}{(\delta x_a)} = 0.
\label{eq:elperturbed}
\end{equation}
where
\begin{eqnarray}
\pder{L}{{\rm v}_a} & = & m_a {\rm v}_a \\
\pder{L}{(\delta x_a)} & = &
-m_a\smb \left( \frac{P_a}{\rho_a^2}+\frac{P_b}{\rho_b^2}\right) \delta x_{ab}
\pder{^2 W_{bc}}{x^2_a}
\nonumber \\
& & - m_a \smb \left[ \left(c_s^2 - \frac{2P_b}{\rho_b} \right)
\frac{\delta\rho_a}{\rho_a^2} +
\left(c_s^2 - \frac{2P_b}{\rho_b} \right) \frac{\delta\rho_b}{\rho_b^2} \right]
\pder{W_{ab}}{x_a}
\end{eqnarray}
giving the SPH form of the linearised momentum equation
\begin{eqnarray} 
\frac{d^2\delta x_a}{dt^2} & = &
-\smb \left(\frac{P_a}{\rho_a^2}+\frac{P_b}{\rho_b^2} \right)\delta x_{ab}
\pder{^2 W_{bc}}{x^2_a} \nonumber \\
& & - \smb \left[ \left(c_s^2 - \frac{2P_b}{\rho_b} \right)
\frac{\delta\rho_a}{\rho_a^2} +
\left(c_s^2 - \frac{2P_b}{\rho_b} \right) \frac{\delta\rho_b}{\rho_b^2} \right]
\pder{W_{ab}}{x_a} \label{eq:perturbed}
\end{eqnarray}
 Equation (\ref{eq:perturbed}) may also be obtained by a direct perturbation of the SPH
equations of motion derived in \S\ref{sec:sphmom}. For linear waves the perturbations are assumed
to be of the form
\begin{eqnarray}
x & = & x_0 + \delta x, \\
\rho & = & \rho_0 + \delta \rho, \\
P & = & P_0 + \delta P.
\end{eqnarray}
where
\begin{eqnarray}
\delta x_a & = & X e^{i(k x_a - \omega t)}, \\
\delta \rho_a & = & D e^{i(k x_a - \omega t)}, \\
\delta P_a & = & c_s^2 \delta \rho_a.
\end{eqnarray}
Assuming equal mass particles, the momentum equation (\ref{eq:perturbed})
becomes
\begin{equation}
-\omega^2 X = -\frac{2m P_0}{\rho_0^2}X \sum_b \left[1 - e^{ik (x_b-x_a)}
\right] \pder{^2 W}{x_a^2} - \frac{m}{\rho_0^2}\left( c_s^2 -
\frac{2P_b}{\rho_b}\right) D \sum_b \left[1 + e^{ik (x_b-x_a)} \right]
\pder{W}{x_a} \label{eq:nearlydispersion}
\end{equation}
From the continuity equation (\ref{eq:sphcty}) the amplitude $D$ of the density
perturbation is given in terms of the particle co-ordinates by
\begin{equation}
D = X m \sum_b \left[1 - e^{ik (x_b-x_a)}\right] \pder{W}{x_a}
\end{equation}
Finally, plugging this into (\ref{eq:nearlydispersion}) and taking the real
component, the SPH dispersion relation (for any equation of state) is given by
\begin{eqnarray}
\omega^2_a & = & \frac{2m P_0}{\rho_0^2} \sum_b [1 - \cos{k (x_a - x_b)}]\pder{^2
W}{x^2}(x_a - x_b, h) \nonumber \\
& & + \frac{m^2}{\rho_0^2}\left(c_s^2 -  \frac{2P_0}{\rho_0}\right) \left[ \sum_b
\sin{k (x_a - x_b)}\pder{W}{x}(x_a - x_b, h) \right]^2,
\end{eqnarray}
For an isothermal equation of state this can be simplified further by setting
$c_s^2 = P_0/\rho_0$. An adiabatic equation of state corresponds to setting
$c_s^2 = \gamma P_0/\rho_0$.

\include{c4app}


\bibliographystyle{klunamed}
\bibliography{jets,starformation,sph,mhd,nsns}

\end{document}